\documentclass[journal]{IEEEtran}
\usepackage{amsmath,amsfonts}
\usepackage{algorithm}
\PassOptionsToPackage{noend}{algpseudocode}
\usepackage{algpseudocode}

\usepackage{array}
\usepackage[caption=false,font=normalsize,labelfont=sf,textfont=sf]{subfig}
\usepackage{textcomp}
\usepackage{stfloats}
\usepackage{url}
\usepackage{verbatim}
\usepackage{graphicx}
\usepackage{cite}
\usepackage{booktabs} 
\usepackage{tabularx} 
\usepackage{multirow}
\usepackage{color}
\usepackage{soul}
\usepackage{pgfplots}
\usepackage{pgfplotstable}
\usepgfplotslibrary{groupplots}
\pgfplotsset{compat=1.18}

\begin{document}

\title{Towards Context-Aware Edge-Cloud Continuum Orchestration for Multi-user XR Services}

\author{Inhar Yeregui, Ángel Martín, Mikel Zorrilla, Roberto Viola, \\ Jasone Astorga, Eduardo Jacob~\IEEEmembership{Senior Member,~IEEE}
\thanks{This work has been submitted to the IEEE for possible publication. Copyright may be transferred without notice, after which this version may no longer be accessible.}
}
\markboth{Journal of \LaTeX\ Class Files,~Vol.~14, No.~8, August~2021}%
{Shell \MakeLowercase{\textit{et al.}}: A Sample Article Using IEEEtran.cls for IEEE Journals}

\IEEEpubid{0000--0000/00\$00.00~\copyright~2026 IEEE}

\maketitle

\begin{abstract}
The growth of multi-user eXtended Reality (XR) applications, spanning fields such as entertainment, education, and telemedicine, demands seamless, immersive experiences for users interacting within shared, distributed environments. Delivering such latency-sensitive experiences involves considerable challenges in orchestrating network, computing, and service resources. Although existing Edge-Cloud Continuum (ECC) solutions provide general-purpose capabilities such as monitoring, scaling, migration, and QoS/QoE assessment, they lack a configurable representation of the internal characteristics of multi-user XR experiences that can directly inform orchestration decisions. This challenge is amplified by the need for high-performance, low-latency connectivity, where 5G and 6G networks provide essential infrastructure to meet the requirements of XR services at scale. This article proposes a context-aware model that parametrises multi-user XR services across four layers of the standard virtualisation architecture and integrates them into a mathematical orchestration framework. The contributions include an analysis of current ECC limitations, a layered formal model, and a proof-of-concept validation. QoS-centric and Cost-centric policies illustrate how the model adapts to different operational priorities. In addition, a controlled ablation using the same orchestration process with and without XR-specific user differentiation shows that application-layer context can influence placement decisions and help preserve the QoS of prioritised users in the evaluated scenario.

\end{abstract}


\section{Introduction}
\IEEEPARstart{e}{Xtended} Reality (XR), encompassing Virtual Reality (VR) and Augmented Reality (AR), has emerged as a transformative technology that seamlessly blends physical and virtual worlds across diverse fields such as entertainment, education, healthcare, and industrial training. By providing deeply immersive and interactive environments, XR enables innovative ways for users to engage with digital content. Within this landscape, multi-user XR services are particularly significant, as they facilitate collaborative, shared experiences among geographically dispersed participants. These services have become essential in a wide range of settings, from critical to social, including medical or aeronautical training \cite{faizan2025integration}, multi-player gaming \cite{vilar2025extended}, remote expert assistance in Industry 4.0 \cite{11385702}, and immersive business meetings \cite{standaert2022business}.

As XR applications grow more sophisticated, their demands on computational, networking, and service-level resources have increased. To ensure seamless experiences through real-time synchronisation and low-latency communication, efficient operation often involves offloading intensive tasks, such as rendering, data processing, and context synchronisation, from end-user equipments (UEs) to more powerful edge or cloud resources \cite{9894514}. This offloading strategy is crucial for XR, as it reduces the burden on UEs that typically lack the processing capacity for high-demand tasks, thereby maintaining real-time performance. Ultimately, as noted in \cite{lagen2023qos}, \cite{luo2023overview}, and \cite{chepkoech2025telementoring}, the integration of 5G and 6G technologies is key to meeting these demands, offering the ultra-low latency, high bandwidth, and massive connectivity required to ensure immersive experiences regardless of user location.

To manage these distributed workloads, the concept of the Edge-Cloud Continuum (ECC) has emerged, which dynamically orchestrates computational tasks across a multi-tiered infrastructure ranging from UEs to edge and cloud nodes. As defined by \cite{moreschini2022cloud}, it addresses two main aspects: (1) extending the traditional cloud to entities like Edge, Fog, and IoT that offer analysis, processing, storage, and data generation \cite{sharma2025review, rahmani2025optimizing}; and (2) enabling flexible resource allocation, placing tasks closer to data sources or users when low latency is needed, while leveraging cloud scalability for less time-sensitive operations \cite{taleb2025survey, hosseinzadeh2025sdn}. Moreschini et al. \cite{moreschini2022cloud} formalise it as \textit{the mechanisms to dynamically orchestrate computation tasks across a continuum of resources available through the network}. This paradigm is primarily enabled by service virtualisation \cite{bhardwaj2021virtualization}, where lightweight containers and Virtual Network Functions (VNFs) \cite{yue2021resource} provide the modularity and rapid instantiation required for the dynamic nature of XR services. The effective operation of the ECC relies on advanced orchestration frameworks that handle the deployment and lifecycle of these virtualised services in real time. Through automated decision-making loops, these frameworks adapt to shifting network conditions and computational loads to improve performance and cost-efficiency. However, while such mechanisms are robust for general applications, XR orchestration presents unique challenges, especially within the shared and distributed environments of multi-user services.

\IEEEpubidadjcol

Multi-user XR services introduce distinct requirements compared to traditional single-user applications, as they involve multiple participants interacting simultaneously in shared virtual or augmented spaces \cite{9894044}. The core challenge lies in ensuring high synchronisation, interactivity, and immersion for all users, regardless of their location, device, or network conditions \cite{9894044}. Real-time interaction management demands that user actions, such as movement or object manipulation, be propagated quickly to maintain context consistency through precise data synchronisation \cite{9132684}. This is further complicated by the need for spatial consistency, where geographically dispersed participants using heterogeneous hardware must perceive a seamless, cohesive environment in real time.

Beyond synchronisation, resource allocation presents a major challenge due to the intensive computation and hardware acceleration required for tasks like remote rendering \cite{11165653}. While edge or cloud servers can alleviate the burden on UEs, managing increasing user counts requires specialised strategies, such as Multipoint Control Units (MCUs) to aggregate inputs and distribute optimised outputs \cite{cernigliaro2020pc}. The orchestration of these sessions is inherently dynamic; computational loads fluctuate with user count, environment complexity, and interaction intensity, often necessitating workload migration across edge nodes to meet latency demands \cite{9132684}. Furthermore, device heterogeneity adds a final layer of complexity, as diverse hardware, ranging from high-end VR headsets to mobile phones, requires adaptive offloading and variable rendering techniques to ensure a consistent and immersive experience for every participant \cite{ko2021functional}.

The convergence of these factors motivates the need for a context-aware ECC orchestration capable of addressing the dynamic nature of multi-user XR services. Existing orchestration frameworks can already react to operational context, including network conditions, resource availability, device capabilities, and user location \cite{sofia2023dynamic}. However, multi-user XR services also involve application-specific characteristics, such as user roles, interaction patterns, perception modes, content properties, rendering quality, and consistency requirements. These characteristics determine how users and service components are affected by infrastructure conditions and cannot be adequately represented by a one-size-fits-all strategy that treats all users and flows uniformly. Among the analysed approaches, we did not identify a unified and configurable framework that explicitly models these internal XR characteristics and their interdependencies and uses them to inform placement and rescheduling decisions across the ECC. This gap motivates the research questions (RQs) addressed in this paper:

\begin{itemize}
    \item RQ1: To what extent do current Edge-Cloud Continuum orchestration solutions incorporate configurable multi-user XR application context into their decisions, and what specific capabilities remain missing?
    \item RQ2: What are the key parameters that characterise the context of a multi-user XR service across the continuum's architectural layers, and how can they be formally modelled to capture their functional interdependencies?
    \item RQ3: How can this layered context parametrisation be incorporated into a multi-objective orchestration model to inform placement and rescheduling decisions while balancing QoS, deployment cost, and operational overhead?
\end{itemize}

To help address the RQs outlined above, this paper presents the following contributions:
\begin{itemize}
    \item C1: A structured analysis of related work and standardisation efforts in ECC orchestration, identifying the specific research gap regarding context-aware support for multi-user XR (related to RQ1).
    \item C2: A formal, layered parametrisation of the XR service context across the ETSI NFV architecture. We introduce the User Engagement Level (UEL) and User Offloading Level (UOL) as analytical abstractions to bridge high-level service requirements with low-level orchestration decisions (related to RQ2).
    \item C3: A multi-objective ECC orchestration model that integrates the layered XR context into placement and rescheduling decisions while balancing Quality of Service (QoS), deployment cost, and operational overhead. A proof-of-concept simulation illustrates policy configurability, and a controlled ablation isolates the effect of XR-specific user differentiation (related to RQ3).
   
\end{itemize}

The remainder of the paper systematically addresses key aspects of the research. Section \ref{relatedwork} answers RQ1 by reviewing the state of the art in ECC orchestration, covering research approaches, standardisation, and XR parametrisation strategies, and identifying the research gap (C1). Section \ref{context} addresses RQ2 by introducing the proposed context formalisation and its application across different layers (C2). RQ3 is covered in Sections \ref{model} and \ref{validation}, which present the mathematical model and its validation, respectively (C3). Finally, Section \ref{conclusions} summarises the findings and outlines directions for future work.

\section{Related Work and needs beyond the State-of-the-Art}
\label{relatedwork}
This section explores related work across three key areas: i) the relevant standardisation efforts in XR and ECC fields, ii) current parametrisation strategies for XR services, and iii) research approaches for ECC orchestration. Together, these topics lay the groundwork for identifying existing gaps towards context-aware ECC orchestration for multi-user XR services. 

\subsection{Standardisation efforts in XR and the ECC}
\label{standardisation}
In the context of standards, XR services are expected to have a salient presence in future mobile networks, driving the evolution of 3rd Generation Partnership Project (3GPP) specifications. Release 18 introduced specialised features such as XR awareness in the Radio Access Network (RAN), enhanced capacity, and energy-saving mechanisms for user equipment \cite{rel18}, building upon the foundations laid in Release 17 \cite{rel17}. Furthermore, 3GPP has defined application layer interfaces specifically for Edge Computing to enable the discovery and administration of Edge Applications, thereby enhancing the performance of 5G-XR services \cite{stafidas2024survey},\cite{christopoulou20255g}.

Interoperability remains a key challenge when integrating the heterogeneous infrastructures and vendors inherent in the ECC. To address this, ISO/IEC 19941:2017 \cite{isointer} outlines four core interoperability aspects:  syntactic, ensuring data formats are correctly interpreted; semantic, ensuring data meaning is preserved across domains; behavioural, ensuring consistent application behaviour across environments; and policy interoperability, which ensures compliance with legal and operational rules. These standards are essential for seamless application migration, which must preserve metadata, dependencies, and runtime environments to maintain functionality. By recommending standardised interfaces and containerisation, this framework supports dynamic migration strategies that minimise service interruptions, a critical requirement for ECC scenarios where systems must adapt to environmental changes in real time.

In parallel, closed-loop automation has emerged as a critical mechanism for balancing the performance of distributed resources in edge-cloud platforms. The European Telecommunications Standards Institute (ETSI) GR ZSM 009-3 report \cite{etsiloop} emphasises that orchestrating these loops must consider the interdependent requirements of individual XR participants, specifically regarding QoS and Quality of Experience (QoE). This approach requires multi-dimensional analytics to balance computational resources with energy efficiency and cost-effectiveness. In this regard, the ITU-T L.14 series provides essential methodologies for assessing the Greenhouse gas (GHG) emissions associated with these communication infrastructures \cite{itutl}. 

\subsection{Parametrisation strategies for XR Services}

Parametrisation is a cornerstone for enabling efficient and scalable XR services, particularly in multi-user environments where interactions occur across distributed systems. Several studies have focused on identifying key parameters that influence XR performance from different network and application perspectives. For instance, Mallik et al. \cite{mallik2024performance} proposed a modelling framework for edge-assisted wireless networks, prioritising parameters such as latency, energy consumption, and data freshness. Similarly, works in \cite{rel17} and \cite{gapeyenko2024overview} explore the integration of XR into 3GPP New Radio (NR) by analysing traffic models and Key Performance Indicators (KPIs), while \cite{morin2023extended} introduced a machine learning-based framework to infer Key Quality Indicators (KQIs) using network-level feature engineering. Beyond network metrics, adaptive rendering techniques have also been explored to dynamically adjust application performance based on user tasks and environmental conditions \cite{liubogoshchev2021adaptive}.

Despite these advancements, existing literature often treats parametrisation in isolation, focusing on specific technical domains like bandwidth or rendering optimisation without addressing the interdependencies across the broader XR ecosystem. A notable challenge lies in the dynamic and heterogeneous nature of multi-user XR, which requires the simultaneous consideration of user interactions, hardware capabilities, and varying network states. While some context-aware frameworks for resource management have been proposed \cite{11126549}, \cite{9184009}, and \cite{chi2025h}, they are frequently constrained to specific use cases or rely on static parameter sets that fail to account for the real-time variability of ECC. This lack of a unified, multi-dimensional model that captures the interplay between multiple users, diverse devices, and distributed infrastructures represents a significant gap. A unified and configurable model capable of representing the internal characteristics of a multi-user XR experience and connecting them with distributed infrastructure decisions remains insufficiently explored.

\subsection{Research approaches for ECC orchestration}

Recent research has begun to incorporate XR-domain information into edge-cloud resource-management decisions. Huang et al. \cite{11318562} use multimodal user intents to guide multi-agent orchestration across communication and computing resources. Santos et al. \cite{11020589} propose a perception-aware approach based on a hybrid QoE model combining network metrics, physiological indicators, and subjective feedback. Ayepah-Mensah et al. \cite{ayepah2026ai} introduce an AI-native cloud-edge framework that uses real-time telemetry, network topology, and microservice dependencies to guide placement and routing for Metaverse workloads. Together, these works illustrate how different contextual abstractions, user intent, perceived QoE, and task dependencies, support different orchestration objectives. 

\begin{table*}[ht!]
\caption{Comparison of general ECC orchestration capabilities in the analysed European projects.}
\centering
\renewcommand{\arraystretch}{1.9} 
\setlength{\tabcolsep}{4pt} 
\label{tab:soa1}
\begin{tabularx}{\textwidth}{|X|c|c|c|c|c|c|c|c|c|c|c|}
\cline{2-12}
\multicolumn{1}{c|}{}
& 
\rotatebox{90}{\textbf{Cognit}\cite{cognit_project}} & 
\rotatebox{90}{\textbf{6G-BRICKS}\cite{6g_bricks_project}} & 
\rotatebox{90}{\textbf{CloudSkin}\cite{cloudskin_project}} & 
\rotatebox{90}{\textbf{CHARITY}\cite{charity_project}} & 
\rotatebox{90}{\textbf{6G-XR}\cite{6g_xr_project}} & 
\rotatebox{90}{\textbf{ACCORDION}\cite{accordion_project}} & 
\rotatebox{90}{\textbf{aeROS}\cite{aeros_project}} & 
\rotatebox{90}{\textbf{INCODE}\cite{incode_project}} & 
\rotatebox{90}{\textbf{TaRDIS}\cite{tardis_project}} & 
\rotatebox{90}{\textbf{OASEES}\cite{oasees_project}} & 
\rotatebox{90}{\textbf{SMARTEDGE}\cite{smartedge_project}} \\ 
\hline
Provide a catalogue of available resources. & \checkmark & \checkmark & \checkmark & \checkmark & \checkmark & \checkmark & \checkmark & \checkmark & X & \checkmark  & \checkmark \\ \hline
Multi-layer real-time monitoring: computational resources, network, applications, etc. & \checkmark & \checkmark & \checkmark & \checkmark & \checkmark & \checkmark & \checkmark & \checkmark & \checkmark & \checkmark  & \checkmark \\ \hline
Incorporates configurable multi-user XR application semantics into placement or rescheduling decisions. & X & X & X & X & X & X & X & X & X & X & X \\ \hline
Continuously evaluates operational service context (e.g., QoS/QoE) to select the appropriate policy or action. & \checkmark & X & \checkmark & \checkmark & X & \checkmark & X & \checkmark & \checkmark & X & \checkmark \\ \hline
Orchestrator able to scale resources for services without interruption. & \checkmark & \checkmark & \checkmark & X & X & \checkmark & \checkmark & X & X & \checkmark & X \\ \hline
Orchestrator able to perform service migrations between nodes without interruption. & \checkmark & X & \checkmark & \checkmark & \checkmark & \checkmark & \checkmark & X & X & \checkmark  & X \\ \hline
\end{tabularx}
\end{table*}

To complement these article-level approaches, we analysed European projects funded by Horizon 2020, Horizon Europe, and the SNS JU that provide implementation-oriented architectures and general operational capabilities for ECC orchestration. These projects address dynamic and resource-intensive services through mechanisms such as AI-driven resource optimisation, multi-layer monitoring, federation, scaling, and migration. Their real-world focus and alignment with standardisation efforts make them relevant for assessing the operational foundations available to support multi-user XR services

The Cognit project \cite{cognit_project} develops cognitive resource management solutions for the ECC through an AI-driven orchestration framework that uses real-time data to optimise resource allocation. It integrates cognitive computing to enhance decision-making in distributed environments, ensuring low-latency, reliable service. A Cloud-Edge Manager handles infrastructure management, offering a resource catalogue, UE offloading, monitoring, auto-scaling, migration, and access control functionalities.

6G-BRICKS \cite{6g_bricks_project} explores next-generation orchestration for distributed systems in 6G networks, focusing on edge-cloud integration to support real-time applications. Using AI models and network slicing, it proposes a unified orchestration layer for managing diverse workloads. The project addresses XR needs like temporal synchronisation and distributed content delivery. Its two-layer Compute Continuum architecture includes an ECC Manager for high-level orchestration and monitoring, and the \(\pi\)-Edge Platform for intra-cluster operations, featuring automation, maintainability, interoperability, autoscaling, and service migration.

CloudSkin \cite{cloudskin_project} proposes a novel virtualisation and orchestration approach to enhance flexibility and scalability in the ECC. It leverages lightweight containers and VMs for seamless service migration across nodes. Designed for resource-intensive applications, its orchestration mechanisms optimise energy use and ensure resilience. The three-layer architecture (Orchestration, Virtualisation, and Infrastructure) enables autoscaling and migration, driven by real-time infrastructure monitoring.

The CHARITY project \cite{charity_project} targets the orchestration needs of XR applications through a multi-tiered framework combining resource management with AI-driven optimisation. Its architecture promotes adaptability by using edge resources for latency-sensitive tasks and cloud for compute-intensive ones. Focusing on immersive media, it offers key insights into orchestrating temporal synchronisation, content adaptation, and QoS in multi-user XR.

6G-XR \cite{6g_xr_project} explores 6G-driven XR orchestration, focusing on seamless Edge/Cloud integration for immersive applications. It investigates distributed resource orchestration to meet XR’s strict latency and bandwidth demands. Using advanced machine learning, the project dynamically optimises resource allocation while sustaining high user engagement and interaction.

ACCORDION \cite{accordion_project} proposes a decentralised orchestration architecture to support scalable coordination in distributed environments. It aims to reduce service deployment latency and improve resource efficiency via edge-to-cloud collaboration. Its work on resource federation and workload balancing supports multi-user XR scenarios with varying computational demands across locations and devices.

The aeROS project \cite{aeros_project} investigates orchestration for autonomous, distributed systems in the ECC. Its framework prioritises robustness and adaptability, using AI for real-time orchestration. It supports autoscaling and resource migration across networks, guided by policy services defining behaviour, data governance, and cognitive capabilities.

Furthermore, the ECC spans the spectrum from centralised cloud to decentralised edge resources. To address challenges such as sovereignty, data heterogeneity, and decentralised decision-making, swarm computing has emerged as a self-organising, collaborative paradigm for hyper-distributed data management, as envisioned in Data Spaces.

Swarm principles are crucial for decentralised coordination among heterogeneous edge systems, where autonomous orchestrators make collective decisions without central control. This enhances adaptability, scalability, and privacy, key in sovereignty-oriented architectures. INCODE \cite{incode_project} develops an open, cloud-native platform for distributed private edge infrastructures. TaRDIS \cite{tardis_project} introduces a language-independent, event-driven model for specifying collaborative node behaviour. OASEES \cite{oasees_project} offers a swarm programmability framework with privacy-preserving Object ID federation, HW accelerators, and Quantum Computing. SMARTEDGE \cite{smartedge_project} enables real-time semantic integration, discoverability, and composability.

The ECC is also critical to future 6G mobile networks, where key research topics include zero-touch management \cite{zts}, complexity simplification \cite{simple}, Service-Based Architectures (SBA), microservices, DevOps, east/west APIs \cite{non}, hardware (HW) accelerators, and serverless computing \cite{serverless}. These are especially relevant for multi-user XR orchestration.

Zero-touch management automates operations, reducing human intervention, errors, and costs. Complexity simplification supports scalable, cost-effective deployment in heterogeneous, distributed environments. SBAs and microservices enable modular, adaptive service composition, improving resilience and resource use. DevOps practices applied to SBAs automate XR service lifecycles, ensuring rapid deployment and continuous optimisation. East/West APIs ensure interoperability and scalable federation across domains, easing infrastructure onboarding. HW accelerators support real-time XR processing through efficient resource allocation. Serverless computing abstracts infrastructure management, streamlining deployment and enabling seamless scaling or migration across cloud-edge-device systems.

\begin{figure}[t!]
    \centering
    \includegraphics[width=0.45\textwidth]{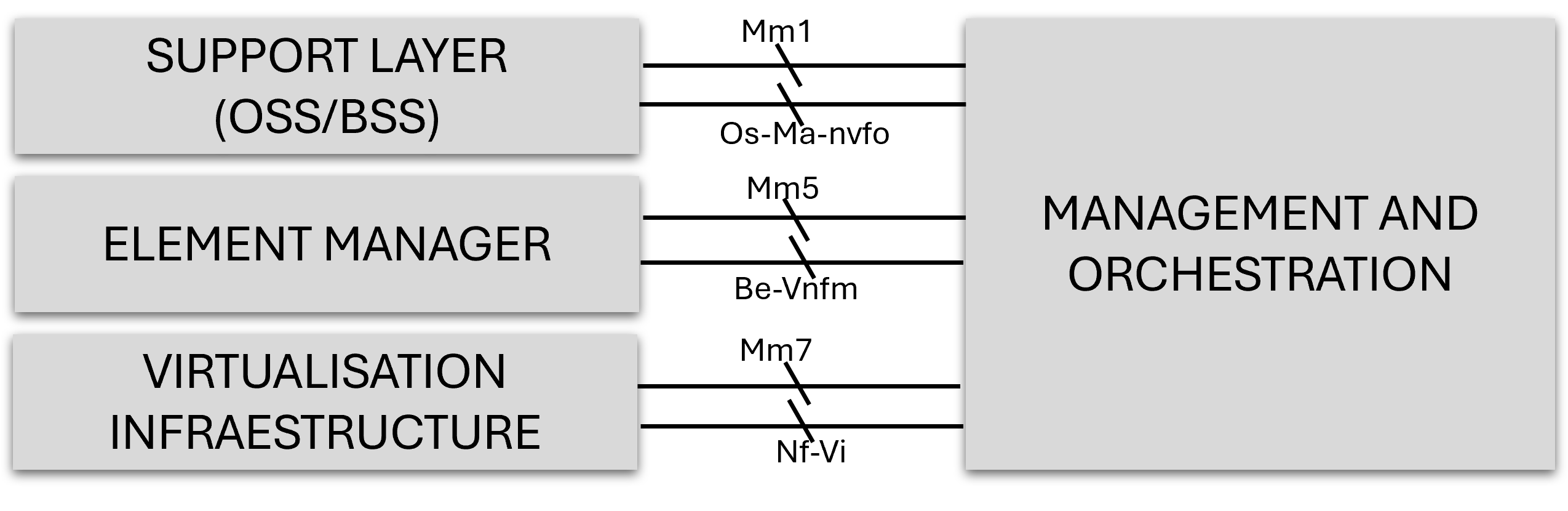}
    \caption{Simplified version of ETSI NFV reference architecture.}
    \label{fig.layers}
\end{figure}

\subsection{Discussion of the identified gap}

The preceding analysis highlights two complementary developments in ECC orchestration. First, the European projects summarised in Table \ref{tab:soa1}, together with relevant standards, provide general capabilities for distributed service management, including resource discovery, monitoring, QoS/QoE-driven adaptation, scaling, migration, interoperability, and automation.

Second, recent XR-oriented research shows that domain-specific information, such as user intent, perceived QoE, and service dependencies, can further improve orchestration decisions. However, these approaches rely on different contextual abstractions tailored to specific objectives, leaving room for a unified and configurable representation of multi-user XR characteristics across architectural layers.

Our work addresses this gap through a layered parametrisation of XR-specific characteristics across four interdependent layers. These parameters are integrated into a multi-objective model to inform placement and rescheduling decisions according to service requirements, infrastructure constraints, deployment cost, and operational overhead.

\begin{figure*}[ht!]
    \centering
    \includegraphics[width=0.98\textwidth]{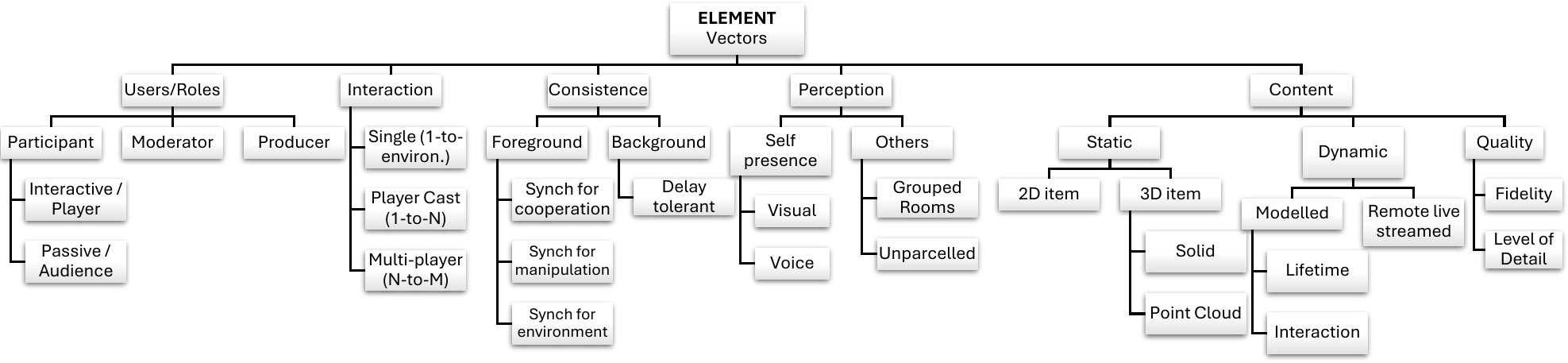}
    \caption{Relevant aspects to consider in the element layer.}
    \label{fig.applayer}
\end{figure*}

\section{Context Modelling and Parametric Decomposition in Multi-user XR Services}
\label{context}

Accurately characterising multi-user XR services within the ECC requires a multi-dimensional understanding of the parameters influencing performance, resource utilisation, and user experience. As established in \cite{sasiain2024preliminary} and \cite{10579130}, different architectural tiers come into play when orchestrating services over 5G and 6G network edge and cloud architectures. Our model follows the ETSI NFV reference architecture shown in Figure \ref{fig.layers}, mapping XR-specific requirements across four interdependent layers. The \textit{element manager layer} manages user-level semantics, such as interactivity and roles, which define the service's immersive nature. These requirements shape the demand on the \textit{virtualisation infrastructure layer}, which governs the execution environment (VCPUs, GPUs, and networking) for offloaded tasks like remote rendering. To ensure long-term viability, the \textit{management and orchestration layer} regulates dynamic resource allocation and migration policies based on real-time QoS metrics (e.g., latency, jitter). Finally, the \textit{support layer} aligns these technical operations with business objectives, managing the trade-offs between Capital and Operational Expenditure (CAPEX/OPEX). 

In the following subsections, we decompose the relevant parameters within their respective layers, as rendered in Figures \ref{fig.applayer}, \ref{fig.computinglayer}, \ref{fig.supportlayer}, and \ref{fig.orchlayer}. While the decomposition presented is both deep and broad, it is not exhaustive. Accordingly, the proposed model could be extended, through more parameters or further constraints, for the inclusion of additional perspectives or emerging concerns that may become relevant as XR services continue to evolve. Thus, the orchestration framework can be adapted to the future mutation and diversification of XR service requirements.

\subsection{Element Manager Layer}

The \textit{element manager} layer manages the lifecycle and operational logic of XR applications deployed within the virtualised infrastructure. As illustrated in Figure \ref{fig.applayer}, this layer defines the service's semantics through several key vectors that directly influence orchestration decisions. First, users are assigned specific roles ($R$), such as interactive participants, moderators, or producers, allowing role-specific QoS priorities to be configured according to the service policy or SLA. The nature of engagement is further defined by interaction types ($I$), which categorise the environment from single-player engagements (1-to-environment) to complex multi-user (N-to-M) relationships.

To maintain immersion, the model distinguishes between foreground consistency ($C_f$), requiring real-time synchronisation for cooperative tasks or environmental coherence, and background consistency ($C_b$) for delay-tolerant processes. Perception is similarly parametrised through self-presence ($P_s$) and the perception of others ($P_o$), supporting clustered (grouped rooms) or fluid (unparcelled) social environments. Finally, the content vector distinguishes between static assets ($D_s$), such as point clouds or solid models, and dynamic elements ($D_d$) that respond to real-time interactions or live streams. These assets are governed by quality parameters ($R_q$), including fidelity and level of detail, which serve as the primary drivers for computational demand and resource allocation in the ECC.

\subsection{Virtualisation Infrastructure Layer}

\begin{figure*}[ht]
    \centering
    \includegraphics[width=0.9\textwidth]{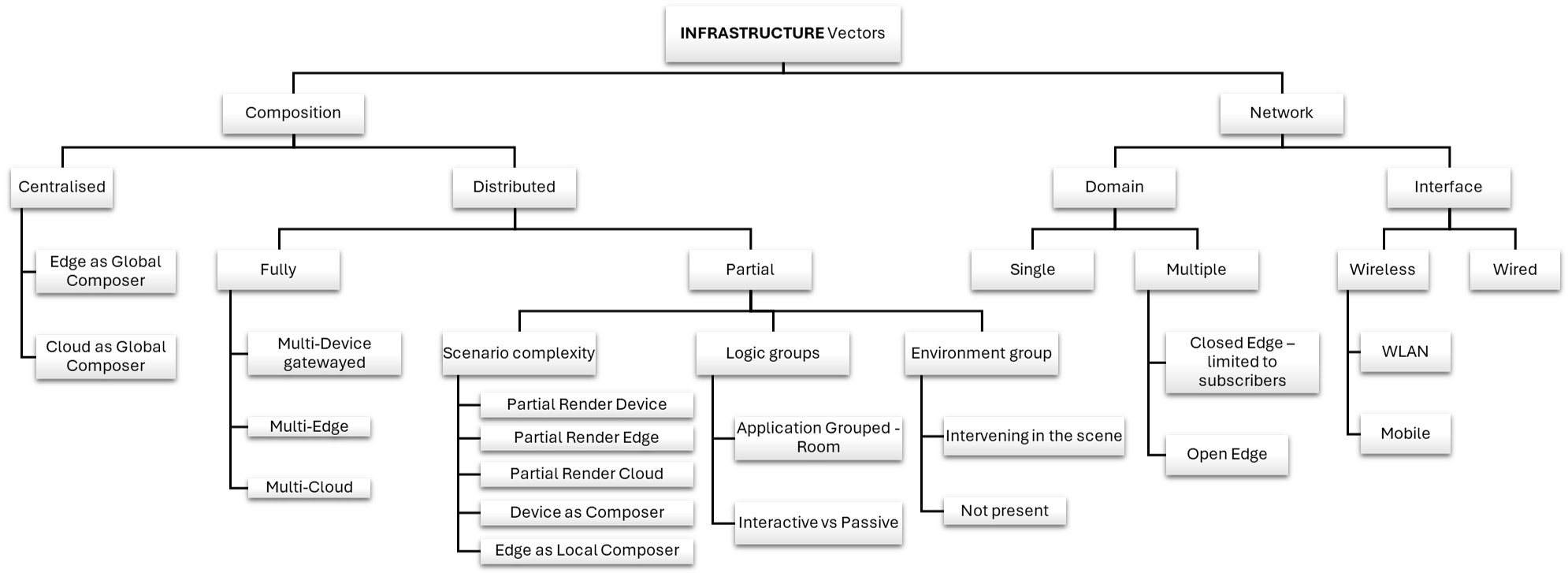}
    \caption{Relevant aspects to consider in the virtualisation infrastructure layer.}
    \label{fig.computinglayer}
\end{figure*}

\begin{figure}[ht]
    \centering
    \includegraphics[width=0.5\textwidth]{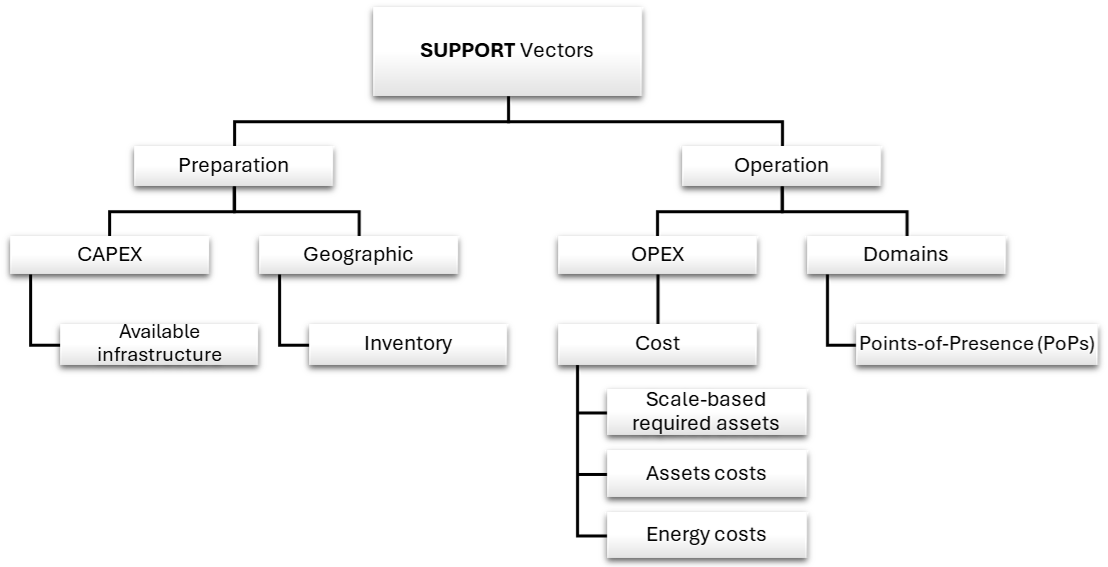}
    \caption{Relevant aspects to consider in the support layer.}
    \label{fig.supportlayer}
\end{figure}

The \textit{virtualisation infrastructure} layer defines the organisation of computing, networking, and storage resources required for service delivery. As shown in Figure \ref{fig.computinglayer}, a primary decision factor is the placement processing mode, which distinguishes between centralised and distributed approaches. While centralised models rely on a single global composer (edge or cloud), distributed models offer higher scalability through fully distributed ($G_d$) setups, leveraging multi-edge or multi-cloud environments, or partially distributed ($G_p$) configurations, where specific tasks like rendering are split between the device and the network. These architectural choices directly impact orchestration complexity, as distributed logic requires more sophisticated coordination to maintain user-level consistency.

From a networking perspective, the model parametrises the connectivity environment to represent transmission and accessibility constraints. Network domains ($N_d$) may comprise single or multiple interconnected domains with different access types ($A_c$), such as open or subscriber-restricted edge nodes. Available interfaces distinguish between wireless ($I_w$) technologies, including WLAN and mobile networks, and wired ($I_l$) infrastructure. Together, these parameters characterise the infrastructure conditions considered when evaluating whether candidate placements can satisfy the XR service requirements and physical resource constraints.

\subsection{Support Layer}

\begin{figure*}[ht!]
    \centering
    \includegraphics[width=0.8\textwidth]{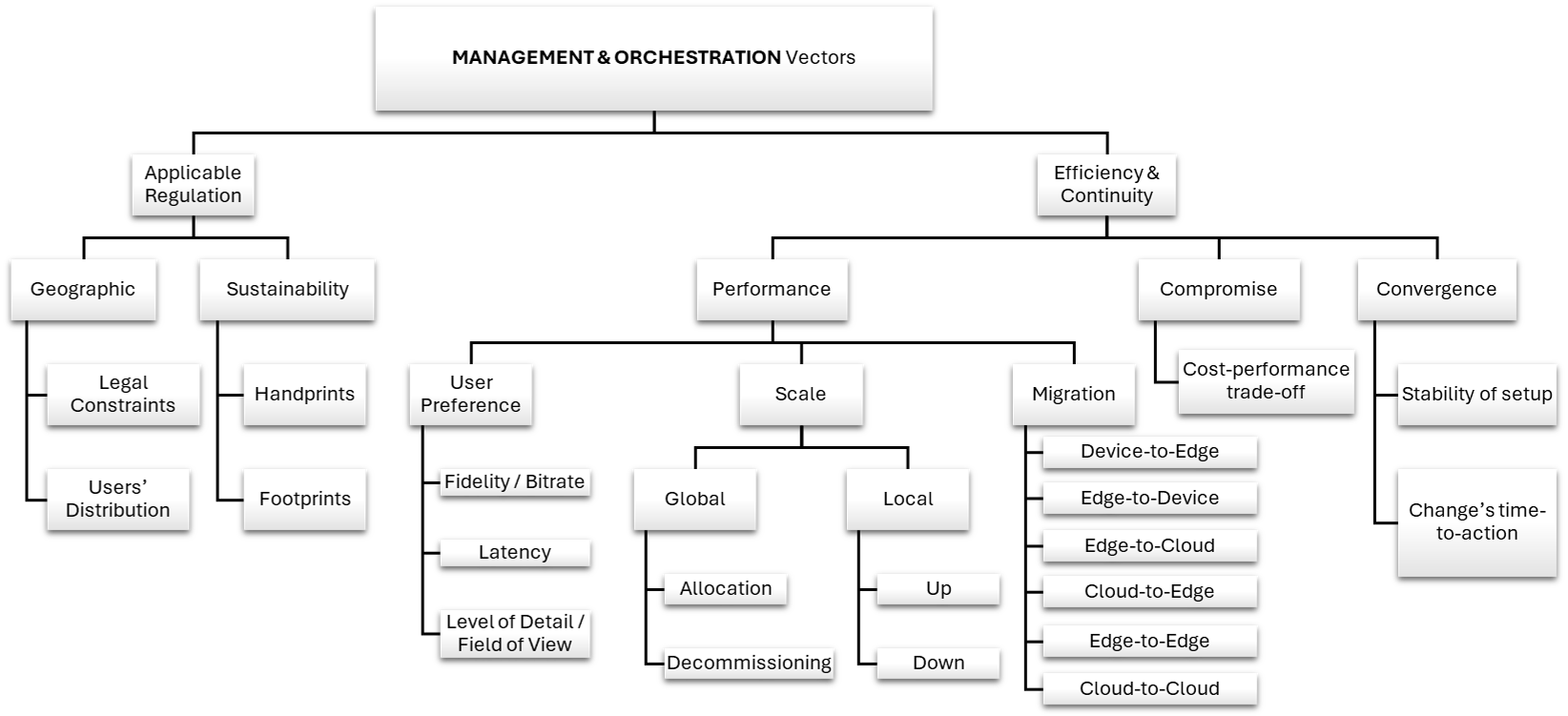}
    \caption{Relevant aspects to consider in the Management and Orchestration layer.}
    \label{fig.orchlayer}
\end{figure*}

The \textit{support} layer captures the business and financial considerations associated with resource preparation and service operation, as depicted in Figure \ref{fig.supportlayer}. During infrastructure preparation, capital expenditure ($C_{capex}$) represents the investment associated with provisioning and scaling resources according to factors such as geographic user distribution and the available asset inventory.

During service operation, operational expenditure ($C_{opex}$) captures the cost implications of resource use. Relevant factors include asset costs, energy consumption ($E_c$), and the availability of Points of Presence ($PoP$) across the network domains involved. These parameters allow placement and resource-allocation decisions to be evaluated not only according to technical requirements, but also in relation to their economic and operational implications.

\subsection{Management and Orchestration Layer}

The \textit{management and orchestration} layer governs the service provisioned to users by regulating system efficiency, continuity, and compliance with operational policies. As illustrated in Fig. \ref{fig.orchlayer}, this layer operates through two primary vectors: Applicable Regulation and Efficiency \& Continuity. The regulatory framework addresses geographic factors ($N_{geo}$), such as regional legal constraints and user distribution, alongside sustainability mandates that balance environmental footprints with positive handprints. This ensures that the infrastructure remains compliant with regional requirements while minimising ecological impact.

To achieve operational excellence, the orchestration process optimises Efficiency \& Continuity by managing performance, trade-offs, and system stability. Performance is maintained by aligning resources with user preferences ($FP_{up}$), which include fidelity, bitrate, and latency requirements. This is achieved through global ($S_g$) or local ($S_l$) scaling mechanisms and seamless migration strategies across the device-to-edge ($M_d$), edge-to-cloud ($M_e$), and inter-node ($M_c$) segments. Furthermore, the system must navigate a hybrid cost-performance compromise ($H_{cp}$), leveraging inputs from the support layer to balance operational expenditure with QoS. Finally, the convergence of these strategies is evaluated through the stability of the setup and the time-to-action ($T_{action}$), ensuring that the service topology adapts proactively to real-world demands without inducing erratic system behaviour.

\section{Problem statement and formulation}
\label{model}

\subsection{Problem description}

\begin{table}[ht!]
\renewcommand{\arraystretch}{1.2} 
\centering
\caption{General parameters of the multi-user XR orchestration model. Parameters marked with * are instantiated in the proof-of-concept validation in Section \ref{validation}.}
\begin{tabularx}{\columnwidth}{|c|X|}
        \hline
        \textbf{Notation} & \textbf{Definition} \\
        \hline
        \multicolumn{2}{|c|}{\textbf{Element Manager Layer}} \\ \hline
        
        $C_f$ & Foreground consistency: synchronisation for cooperation, manipulation, and environment. \\ \hline
        $C_b$ & Background consistency: delay-tolerant processes. \\ \hline
        $D_s$ & Static content: 2D items, 3D items (solid models, point clouds). \\ \hline
        $D_d$ & Dynamic content: Modeled content (defined lifetime, interactive) and Remote live-streamed data. \\ \hline
        $I^{*}$ & Interaction types: Single-player (1-to-environment), Player Cast (1-to-N), Multi-player (N-to-M). \\ \hline
        $U^{*}$ & List of users in the XR experience $<u_1, u_2, ... u_n>$. \\ \hline
        $P_s^{*}$ & Self-perception: Visual or Voice-based presence. \\ \hline
        $P_o$ & Perception of others: Grouped rooms or Unparcelled space. \\ \hline
        $R^{*}$ & User role: participants (interactive or passive), moderators, producers. \\ \hline
        $R_q^{*}$ & Rendering quality of the content in the XR experience. \\ \hline

        \multicolumn{2}{|c|}{\textbf{Virtualisation infrastructure Layer}} \\ \hline
       
        $A_c$ & Access type: Closed Edge (limited to subscribers) or Open Edge. \\ \hline
        $G_d$ & Fully distributed: Multi-device, Multi-edge, Multi-cloud setups. \\ \hline
        $G_p$ & Partially distributed: Scenario-based orchestration (Device rendering, edge rendering, cloud rendering). \\ \hline
        $I_w$ & Wireless interfaces: WLAN, Mobile networks. \\ \hline
        $I_l$ & Wired interfaces: Traditional network infrastructure. \\ \hline
        $N_d$ & Network domain: Single or Multiple. \\ \hline
        $J^{*}$ & Set of placements. \\ \hline
        $N_{pl}^{*}$ & Set of nodes involved in a placement. \\ \hline

        \multicolumn{2}{|c|}{\textbf{Support Layer}} \\ \hline
        $C_{capex}$ & Capital expenditure for infrastructure scaling and provisioning. \\ \hline
        $C_{opex}^{*}$ & Operational expenditure (asset management, energy consumption). \\ \hline
        $E_c$ & Energy cost consumed by resources. \\ \hline
        $PoP^{*}$ & Number of Points of Presence (PoPs) for geographic reach. \\ \hline
        $R_{usage}^{*}$ & Computational and network resource demand of a user. \\ \hline
        $R_{alloc}^{*}$ & Allocated computational and network resources in the system. \\ \hline
        $R_{max}^{*}$ & Total computational and network resources in the system. \\ \hline
        
        \multicolumn{2}{|c|}{\textbf{Management and Orchestration Layer}} \\ \hline
        $FP_{up}^{*}$ & User preferred settings: Fidelity, Bitrate, Level of Detail vs Latency. \\ \hline
        $H_{cp}$ & Hybrid orchestration: Cost-performance trade-offs. \\ \hline
        $M_d$ & Migration: Device-to-Edge, Edge-to-Device. \\ \hline
        $M_e$ & Migration: Edge-to-Cloud, Cloud-to-Edge. \\ \hline
        $M_c$ & Migration: Edge-to-Edge, Cloud-to-Cloud. \\ \hline
        $N_{geo}$ & Number of distinct geographical regions where active XR users are located. \\ \hline        
        $P$ & Placement processing mode: Centralised or Distributed. \\ \hline
        $S_g^{*}$ & Global scalability: Dynamic allocation, resource decommissioning. \\ \hline
        $S_l$ & Local scalability: On-demand up/down scaling. \\ \hline
        $T_{action}^{*}$ & Time-to-action: Ensuring rapid system response. \\ \hline
\end{tabularx}
\label{tab:xr_parameters}
\end{table}
Orchestrating multi-user XR requires dynamic resource allocation and placement across the ECC to balance user QoS, computational demand, deployment cost, and rescheduling overhead. The proposed formulation integrates layered XR application context into candidate-placement evaluation, translating service characteristics into user priorities and resource requirements that guide placement, scaling, and migration decisions. The mathematical notations categorised by layer in Table \ref{tab:xr_parameters}, derived from Figures \ref{fig.applayer}, \ref{fig.computinglayer}, \ref{fig.supportlayer}, and \ref{fig.orchlayer}, provide the basis for this formulation.

\subsection{Parameter abstraction}
\label{parameter_abstraction}

The \textit{element manager} layer parameters play a crucial role in shaping user experiences and influencing system resource demand. However, many of these parameters are qualitative and cannot be directly incorporated into the numerical orchestration process. To bridge this gap, we propose two configurable analytical abstractions: the User Engagement Level (UEL) and the User Offloading Level (UOL). These abstractions translate the multi-dimensional influence of the layer-specific parameters into normalised values that can be incorporated into placement evaluation.

To ensure mathematical consistency across heterogeneous parameters, we define a mapping function $\phi(p) \to [0, 1]$, which translates qualitative attributes such as user roles and interaction types, into normalised numerical values according to the service policy, SLA, or use-case requirements.

    \subsubsection{\textbf{User Engagement Level (UEL)}}
        
    The UEL quantifies the depth of a user’s interaction and immersion under a given service policy. It reflects the intensity of their engagement with the environment, where higher values indicate a deeper involvement that necessitates stricter QoS priority. We define $UEL_u$ as a weighted additive model to allow for compensability and scenario-specific tuning:
    
    \begin{equation} 
    \label{eq:UEL_formal}
        UEL_u = \sum_{k \in \mathcal{K}} \omega_k \cdot \phi(p_{u,k})
    \end{equation}
    
    Where $\mathcal{K} = \{ I_u, R_u, P_{s,u}, P_{o,u}, D_s, D_d, R_{q,u}, C_f, C_b \}$ is the set of parameters from the \textit{element manager} layer, and $\omega_k$ represents the relative importance of each factor (with $\sum \omega_k = 1$). For instance, the interaction type $I_u$ is mapped such that $\phi(I_{multiplayer}) > \phi(I_{singleplayer})$, as collaborative interactions demand higher synchronisation. Similarly, the role $R_u$ prioritises producers or moderators over passive audience members. By using this weighted sum, the UEL becomes a continuous metric in the range $[0, 1]$ that prioritises resource allocation for the most "active" users in the XR session.

    \subsubsection{\textbf{User Offloading Level (UOL)}} 
    
    The UOL is an abstraction that quantifies the extent to which processing and rendering tasks are migrated from the user’s device to the edge or cloud infrastructure. This parameter is essential for the orchestrator to determine the workload distribution across the ECC. Since the infrastructure models $G_d$ (fully distributed) and $G_p$ (partially distributed) are structural choices, we define $UOL$ as the aggregate offloading factor associated with candidate placement $j$:

    \begin{equation}
    \label{eq:UOL_formal}
        UOL_j = \delta \cdot \phi(G_d) + (1-\delta) \cdot \phi(G_p)
    \end{equation}

    Where $\delta \in \{0, 1\}$ is a binary selector that identifies the distribution model of the placement. A higher $UOL_j$ indicates that a larger portion of the rendering pipeline is performed remotely, thereby increasing the network bandwidth requirement while reducing the computational strain on UE. 
    
    In practical deployments, $UEL$ and $UOL$ can be instantiated from XR service descriptors and operator-defined policies. The service provider may define profiles for user roles, interaction modes, perception types, rendering qualities, and consistency requirements, assigning normalised $\phi(\cdot)$ values and parameter weights $\omega_k$ according to the SLA or use-case requirements. For example, an interactive participant in an N-to-M session with point-cloud representation could be assigned a higher $UEL_u$ than a passive audience user, while $UOL_j$ is derived from the selected execution profile associated with candidate placement $j$, such as local, device-edge split, edge, or cloud processing. During runtime, these abstractions contribute to user prioritisation within placement evaluation, while infrastructure-level metrics are incorporated through the constraint indicators and objective-function terms.

\subsection{Definition of the system constraints}
\label{constraints}

The orchestration model is governed by a set of five fundamental constraints that act as a multi-stage filter to define the feasible solution space, spanning technical feasibility (RIC and QoSC), economic viability (PCC), and operational continuity (SOC and MOC).

   \begin{itemize}
       \item\textbf{Resource Infrastructure Constraint (RIC)}: To ensure system stability and avoid computational exhaustion, the resources required by the XR service must not exceed the infrastructure's limits. This constraint is evaluated through two distinct boolean indicators that differentiate between the physical capacity of the node and the current resource assignment. 
       
       First, the Resource Capacity Indicator ($\mathcal{I}_{RC,j}$) represents the physical viability of a placement $j$. It acts as a hard filter by verifying if the total available capacity across the active Points of Presence ($PoP$) can physically support the cumulative user demand. If $\mathcal{I}_{RC,j}$ is \textit{false}, the placement is discarded as it is technically unfeasible.
        \begin{equation}
        \label{eq:irc_boolean}
        \scalebox{0.82}{ 
        $
        \mathcal{I}_{RC,j} =
        \begin{cases}
        \text{true} & \text{if } \sum_{u \in U} R_{usage,u} \cdot \phi(D_s, D_d) \leq R_{max}(PoP) \\
        \text{false} & \text{otherwise}
        \end{cases}
        $
        }
        \end{equation}

        Then, the Resource Allocation Indicator ($\mathcal{I}_{RA,j}$) assesses whether the resources currently reserved or allocated to the service instance (e.g., vCPU/RAM limits in a container) are sufficient for the detected context. This indicator is crucial for the rescheduling phase, as it determines if a simple scaling operation is required even if the node has sufficient physical capacity.
        \begin{equation}
        \label{eq:ira_boolean}
        \scalebox{0.82}{ 
        $
        \mathcal{I}_{RA,j} =
        \begin{cases}
        \text{true} & \text{if } \sum_{u \in U} R_{usage,u} \cdot \phi(D_s, D_d) \leq R_{alloc,j}(PoP) \\
        \text{false} & \text{otherwise}
        \end{cases}
        $
        }
        \end{equation}
        
        Where $R_{usage,u}$ is the demand per user, influenced by the complexity of static ($D_s$) and dynamic ($D_d$) content through the scaling factor $\phi$. $R_{max}(PoP)$ denotes the total hardware limit of the node, while $R_{alloc,j}$ represents the logical resources currently provisioned for the service instance in placement $j$.
       
        \item \textbf{Quality of Service Constraint (QoSC)}: The interactive nature of XR imposes strict constraints on latency ($L_u$) and throughput ($Th_u$). These must satisfy the application-specific thresholds $L_{max_u}$ and $Th_{min_u}$, which are defined by the service type. $L_{max_u}$ represents the maximum tolerable delay (e.g., Motion-to-Photon latency) to avoid motion sickness and maintain interactivity, while $Th_{min_u}$ is the minimum bitrate required for the service. We define the state of this constraint for a placement $j$ through the boolean indicator $\mathcal{I}_{QoSC,j}$:

        \begin{equation}
        \label{eq:qos_boolean}
        \scalebox{0.85}{ 
        $
        \mathcal{I}_{QoSC,j} =
        \begin{cases}
        \text{true} & \text{if } \forall u \in U: 
        \begin{cases}
        L_u \leq L_{max_u}  \\
        Th_u \geq Th_{min_u} 
        \end{cases}
        \\
        \text{false} & \text{otherwise}
        \end{cases}
        $}
        \end{equation}
    
        
        To evaluate this indicator, the end-to-end latency is decomposed into processing and network components, modeled as follows:
        
    
        \begin{equation}
            \label{eq:Lu}
            L_u = L_{proc_u} + L_{net_u}
        \end{equation}
        
        \begin{equation}
        \label{eq:Lproc_sum}
        L_{proc_u} = \sum_{k \in \mathcal{K}_{proc}} \tau_k \cdot \phi(k_u)
        \end{equation}
        \begin{equation}
        \label{eq:Lnet_sum}
        L_{net_u} = d_{int} + \sum_{m \in \mathcal{K}_{net}} \delta_m \cdot \phi(m_u)
        \end{equation}
    
        In Equation \eqref{eq:Lproc_sum}, $\mathcal{K}_{proc}$ denotes the set of element manager parameters ($C_f, C_b, D_s, D_d, I, P_s, P_o, R_q$), where $\tau_k$ represents the processing time (in ms) associated with each factor's complexity. For instance, higher rendering fidelity ($R_q$) or complex dynamic content ($D_d$) linearly increase $L_{proc_u}$ according to the node's capacity.
    
        In Equation \eqref{eq:Lnet_sum}, network delay is decomposed into the interface-specific delay $d_{int}$ (determined by $I_w$ or $I_l$) and the routing/synchronisation overhead defined by the set $\mathcal{K}_{net} = \{ N_d, A_c, PoP, N_{geo} \}$. The coefficients $\delta_m$ capture the impact of traversing multiple domains ($N_d$) or synchronising dispersed users ($N_{geo}$) across service instances.
    
        Similarly, the throughput $Th_u$ is formulated as the base bitrate required by the content, scaled by the rendering and infrastructure configuration:
        \begin{equation}
        \label{eq:Th_prod}
        Th_u = \left( BR_{base} \cdot \phi(D_s, D_d, |U|, R) \right) \cdot \mu(R_q, G_p)
        \end{equation}
    
        Where $BR_{base}$ is the standard bitrate determined by the content type and the number of active users $|U|$. This base demand is scaled by a factor $\mu \geq 1$ depending on the rendering quality ($R_q$) and the offloading scheme ($G_p$), reflecting that remote rendering increases bandwidth demand significantly compared to local processing.


    \item \textbf{Placement Cost Constraint (PCC)}: The cost constraint ensures that the total operational expenditure of deploying the XR service within placement $j$ does not exceed the predefined budget $C_{opex}$. We define the state of this constraint through the boolean indicator $\mathcal{I}_{PCC,j}$:
    
    \begin{equation}
    \label{eq:pcc_boolean}
    \scalebox{1}{ 
    $
    \mathcal{I}_{PCC,j} =
    \begin{cases}
    \text{true} & \text{if } Cost_{PL_{j}} \leq C_{opex} \\
    \text{false} & \text{otherwise}
    \end{cases}
    $}
    \end{equation}
    
    The cost of a placement \textit{j} is calculated by summing the costs of all the nodes involved in that placement $i \in N_{pl}$:.

    \begin{equation}
    \label{eq:cost_total}
    Cost_{PL{j}} = \sum_{{i \in N_{\text{pl}}}} Cost_{i}
    \end{equation}
    
    For each node $i$, the cost is decomposed into computational and energy-related expenses:
    
    \begin{equation} 
    \label{eq:costi}
    Cost_{i} = \varrho_i \cdot \sum_{u \in U} R_{usage_u} + \xi_i \cdot E_{c_i}
    \end{equation}

    Where $\varrho_i$ represents the unit cost per computational resource (e.g., currency per vCPU/RAM unit) for node $i$, while $R_{usage_u}$ denotes the resource demand of user $u$. The second term captures the energy cost, where $E_{c_i}$ is the energy consumed by the node's resources and $\xi_i$ is the applicable energy tariff (e.g., currency per kWh).


    \item \textbf{Scalability Overhead Constraint (SOC)}: The scalability overhead constraint ensures that the system can dynamically expand or contract resources to accommodate shifting service demands without introducing excessive instability. We define the state of this constraint through the boolean indicator $\mathcal{I}_{SOC,j}$, which evaluates a placement based on two criteria: the physical availability of resources for scaling and the operational impact of the scaling action. The indicator is defined as follows:

    \begin{equation}
    \label{eq:soc_boolean}
    \scalebox{0.9}{ 
    $
    \mathcal{I}_{SOC,j} =
    \begin{cases}
    \text{true} & \text{if } S_g \geq \Delta R_{usage} \text{ and } S_{OH_j} \leq S_{OH_{max}} \\
    \text{false} & \text{otherwise}
    \end{cases}
    $}
    \end{equation}

    Where $\Delta R_{usage}$ represents the required resource increment, and $S_{OH_{max}}$ is the maximum allowable overhead. To evaluate the physical feasibility, we define the aggregate scalability potential ($S_g$) as the total capacity available for resource adjustment across the nodes hosting the service:
   
    \begin{equation}
    \label{eq:sg_formal}
    S_g = \sum_{{i \in N_{pl}}} S_{l,i}
    \end{equation} 
    where  \( S_{l,i} = (R_{max_i} - R_{alloc_i})\) if node \( i \) can scale up resources, else \( 0 \). 

    Furthermore, to ensure service stability, the overhead incurred by the scaling operation ($S_{OH_j}$) is modeled as a function of the actuation time ($T_{action, j}$) relative to the service's sensitivity:
    
   \begin{equation}
   \label{eq:soh_formal}
    S_{OH_j} = \eta \cdot T_{action, j} 
    \end{equation}
    
   In Equation \eqref{eq:soh_formal}, $\eta$ is a weighting coefficient that represents the sensitivity of the XR service to scaling-induced jitter.
    
   \item\textbf{Migration Overhead Constraint (MOC)}: Service migration between nodes introduces overhead in the form of state synchronisation, connection handovers, and potential downtime. The migration overhead constraint ensures that these disruptions do not exceed a predefined maximum threshold $M_{OH_{max}}$. We define the state of this constraint through the boolean indicator $\mathcal{I}_{MOC,j}$:

    \begin{equation}
    \label{eq:moc_boolean}
    \scalebox{1}{ 
    $
    \mathcal{I}_{MOC,j} =
    \begin{cases}
    \text{true} & \text{if } M_{OH_j} \leq M_{OH_{max}} \\
    \text{false} & \text{otherwise}
    \end{cases}
    $}
    \end{equation}

    $M_{OH_j}$ is modelled as a function of the total actuation time $T_{action, j}$ required to transfer the service context across different virtualised infrastructures:

    \begin{equation}
    \label{eq:moh_formal}
    M_{OH_j} = \psi \cdot T_{action,j}(M_d, M_e, M_c) + \sigma(P)
    \end{equation}

    In Equation \eqref{eq:moh_formal}, the actuation time depends on whether the migration occurs between device and edge ($M_d$), edge and cloud ($M_e$), or across different nodes within the same layer ($M_c$). The coefficient $\psi$ represents the service's sensitivity to migration-induced downtime, while $\sigma(P)$ is an additional overhead constant that accounts for the coordination delay if the orchestration decision is distributed ($P$) rather than centralised. By relating the overhead directly to physical actuation times and infrastructure types, the model ensures that the system only triggers a migration if the resulting state transition can be completed without jeopardising the stability of the XR experience.
    
    
\end{itemize}

   In the proposed orchestration framework, the aforementioned conditions are formalised as a set of five hard constraints, represented by the binary indicators $\mathcal{I}_{RC}, \mathcal{I}_{QoSC}, \mathcal{I}_{PCC}, \mathcal{I}_{SOC}$, and $\mathcal{I}_{MOC}$. The model enforces a strict feasibility policy where a candidate placement $j$ is considered valid only if all these indicators are satisfied simultaneously. This ensures that any placement selected by the system fundamentally meets the minimum resource, latency, and operational requirements. Furthermore, the resource availability indicator, $\mathcal{I}_{RA}$, serves as a specialised trigger within the decision-making logic; it determines whether the infrastructure can support a vertical scaling action or if a service migration is required. This deterministic structure provides a reliable foundation for the orchestration process, ensuring that the subsequent multi-objective evaluation is only performed on nodes that are technically capable of hosting the XR service.

    \subsection{Model Formulation}
    
    For each feasible placement, the model computes $F_j$ to maximise QoS while minimising placement cost and rescheduling overhead. The three terms are normalised to $[0, 1]$ before aggregation.

    \begin{enumerate}
        \item \textbf{Maximising overall QoS}, \( QoS_{PL{j}} \), which is calculated as the sum of the weighted individual user QoS values for a given placement, allowing the system to prioritise specific roles or engagement levels:

        \begin{equation}
        \label{Qos_pl_j}
        QoS_{PL{j}} = \sum_{u \in U} w_{u,j} \cdot QoS_u 
        \end{equation}        

    \begin{itemize}
        \item \( w_{u,j} \): represents the importance or priority of user $u$ when evaluating candidate placement $j$. We define this weight as a linear combination of the user's engagement level, the offloading level associated with the candidate placement, and the user's preferences:
        \begin{equation}
        \label{eq:wu}
        w_{u,j} = \theta_1 UEL_u + \theta_2 UOL_j + \theta_3 \phi(FP_{up_u})
        \end{equation}
        Where $UEL_u$ represents the user engagement level, $UOL_j$ represents the offloading level associated with candidate placement $j$ and $\phi(FP_{up_u})$ maps user preferences (fidelity vs. latency) to a numerical scale. The coefficients $\theta_k$, with $\sum_k \theta_k = 1$, allow administrators to determine the relative influence of each factor on the resulting user priority.

        \item \( QoS_u \): represents the QoS for a given user in terms of latency and throughput. It reflects how well the system satisfies the user's needs relative to their maximum acceptable latency and minimum viable throughput:
       \begin{equation}
       \label{eq:qos_u}
           QoS_u = \omega_L \cdot \left( 1 - \frac{L_u}{L_{max_u}} \right) + \omega_{Th} \cdot \left( \frac{Th_u - Th_{min_u}}{Th_u} \right)
        \end{equation}
    
    where $\omega_L + \omega_{Th} = 1$. This utility-based formulation ensures that $QoS_u$ remains within the range $[0, 1]$, where 1 represents perfect compliance with the thresholds and 0 indicates a violation of the requirements. Definitions of $L_u$ and $Th_u$ have been described in Equations \eqref{eq:Lu}, \eqref{eq:Lproc_sum}, \eqref{eq:Lnet_sum}, and \eqref{eq:Th_prod}.
    \end{itemize}
    
    To normalise $QoS_{PL_j}$, we treat it as a weighted average of the individual user experiences. Since each $QoS_u$ is bounded between 0 and 1, the maximum value of the aggregate sum for candidate placement $j$ is reached when all users experience ideal conditions, equalling the sum of the assigned weights $\sum w_{u,j}$. Thus, the normalised $QoS_{PL_j}'$ is calculated as:

    \begin{equation}
    \label{eq:QoS_pl_normalised}
    QoS_{PL_j}' = \frac{QoS_{PL{j}}}{\sum_{u \in U} w_{u,j}}
    \end{equation}

    \item \textbf{Minimising the cost of service placement}. It represents the cost of deploying the XR service on placement \( j\), calculated by summing the cost associated with each node of the placement as previously defined in Equations \eqref{eq:cost_total} and \eqref{eq:costi}. This includes the cost of computational resources (CPU, GPU, RAM, etc.) and the energy consumption translated into monetary value through specific tariffs.

    To normalise \( Cost_{PL_j} \), we define a maximum reference cost \( Cost_{PL_j}^{max} \), corresponding to the highest possible resource cost among all nodes in the infrastructure when their total CPU, RAM and GPU capacities are weighted by their respective unit costs. The placement cost of each candidate node is then normalised by dividing it by this reference value.
    
    \begin{equation}
    \label{eq:cost_max_global}
    Cost_{PL_j}^{max} = \max_{i \in N_{infra}} \left( \varrho_i \cdot R_{i}^{max} + \xi_i \cdot E_{c_i}^{max} \right)
    \end{equation}
    
    Where $ R_{i}^{max} $ represents the total capacity (CPU, RAM, GPU) of node $ i $, and $ N_{infra} $ is the set of all available nodes in the infrastructure. The placement cost is then normalised by dividing the cost of the nodes in placement $j$ by this global reference value:    
    
    \begin{equation}
    \label{eq:cost_pl_normalised}
    Cost_{PL_j}' = \frac{ \sum_{i \in N_{pl}} Cost_i }{ |N_{pl}| \cdot Cost_{PL_j}^{max} }
    \end{equation}

    This approach ensures that $ Cost_{PL_j}' $ is correctly scaled between 0 and 1. By using a global infrastructure maximum as a benchmark, the orchestrator can effectively compare the economic efficiency of different placements regardless of their topological location.

    \item \textbf{Minimising the rescheduling operation overhead}. This refers to the operational "penalty" associated with state transitions, such as scaling resources or migrating services between nodes. These operations introduce temporary instabilities including delays, synchronisation tasks, and potential packet loss. The total rescheduling operation overhead for a candidate placement $j$ is expressed as the sum of its scaling and migration penalties:
    \begin{equation}
    \label{eq:oh_ro_j}
    OH_{RO_j} = S_{OH_j} + M_{OH_j}
    \end{equation}

    where \( S_{OH_j} \) and \( M_{OH_j} \) are the scaling and migration overheads defined in Equations \eqref{eq:soh_formal} and \eqref{eq:moh_formal}, respectively.
     
        

    To normalise $OH_{RO_j}$ and ensure it can be integrated into the multi-objective function, we define the maximum allowable overhead $OH_{RO_j}^{max}$. This reference value corresponds to the sum of the maximum operational thresholds established in the system constraints:
    \begin{equation}
    \label{eq:oh_ro_max}
    OH_{RO_j}^{max} = S_{OH_{max}} + M_{OH_{max}}
    \end{equation}
    
    The normalised rescheduling operation overhead, $OH'_{RO_j}$, is then calculated as the ratio between the current transition cost and the total disruption budget:    \begin{equation}
    \label{eq:oh_ro_normalised}
    OH'_{RO_j} = \frac{OH_{RO_j}}{OH_{RO_j}^{max}} =\frac{S_{OH_j} + M_{OH_j}}{S_{OH_{max}} + M_{OH_{max}}}
    \end{equation}

    By using the sum of the thresholds ($S_{OH_{max}}$ and $M_{OH_{max}}$) as the denominator, we ensure that $OH'_{RO_j} \in [0, 1]$ for any feasible placement. A value of 0 represents the baseline state where no rescheduling operations are performed. Conversely, a value of 1 indicates a high-impact transition that consumes the entire disruption budget permitted by the system's stability constraints.
    \end{enumerate}
    
    In order to tune the applicable cost-performance trade-off of hybrid orchestration operations, a vector with weights for each involved dimension, service QoS (\( \alpha \)), service placement costs (\( \beta \)), and operational overheads (\( \gamma \)), is defined as:
    \begin{equation}
    \vec{H_{cp}} = \begin{pmatrix}\alpha & \beta & \gamma\end{pmatrix}
    \end{equation}
    
    With this in mind, our target function is described as follows: 
    \begin{equation}
    \label{eq:Fj}
    \begin{split}
    F_j =   \ \alpha \cdot QoS'_{PL_j} 
    - \beta \cdot Cost'_{PL_j} 
    - \gamma \cdot OH'_{RO_j}
    \end{split}
    \end{equation}
    \begin{equation}
    \label{eq:Fbest}
    \begin{split}
    F_{\text{best}} =  \max_{j \in J} \left(  F_j \right)
    \end{split}
    \end{equation}
The target function evaluates each feasible placement \( j \) through the objective value \(F_j\), which balances overall \( QoS\), placement cost, and rescheduling operation overhead. The maximum objective value across all feasible placements is denoted by \(F_{best}\), and the placement that achieves this value is referenced as \(j_{best}\).


\subsection{State Machine/Graph Model}

\begin{figure}[t]
    \centering
    \includegraphics[width=1\linewidth]{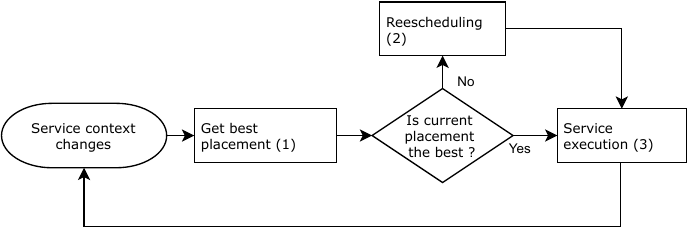}
    \caption{State Machine for context-aware multi-user XR service orchestration.}
    \label{fig.statemachine}
\end{figure}

Figure \ref{fig.statemachine} illustrates the overall workflow of the orchestration model, implemented as an asynchronous, event-driven controller operating at the service-placement timescale. An external telemetry layer continuously monitors the infrastructure and XR service state, while orchestration is triggered only by significant context changes, such as user connections or disconnections, changes in resource availability, or QoS violations. Upon each trigger, the orchestrator evaluates a recent state snapshot to determine whether the current placement should be maintained or rescheduled. Packet- or frame-level fluctuations do not directly trigger placement evaluation, reducing unnecessary re-evaluations and associated rescheduling overhead.

These phases are described in more detail as follows:

\begin{enumerate}

    \item \textbf{Get Best Placement}: To find the best placement, $ F_{best} $, Algorithm \ref{alg:xr_orchestration} filters candidate placements through the five hard constraints, evaluates \(F_j\) for feasible candidates, and selects \(j_{best}\) as the highest-scoring placement.
    
    \begin{algorithm}[ht]
    \renewcommand{\algorithmicrequire}{\textbf{Input:}}
    \renewcommand{\algorithmicensure}{\textbf{Output:}}
    \caption{Get best placement}
    \label{alg:xr_orchestration}
    \begin{algorithmic}
    \Function{FindBestPlacement}{ $U$, $Topology$, $j_{current}$, $R_{usage}$,  $\alpha$, $\beta$, $\gamma$}
    
        \Require $U$ \Comment{Set of active users}
        \Require $Topology$ \Comment{Topology information}
        \Require $j_{current}$ \Comment{Current placement}
        \Require $R_{usage}$ \Comment{Total resource usage}
        \Require $\{\alpha, \beta, \gamma\}$ \Comment{Weight factors}
        \Ensure $j_{best}$ \Comment{Best placement}
    
        \State \textbf{1. Check Constraint Indicators}
        \ForAll{Placements}
            \State 1. Resource Infrastructure Constraint ($\mathcal{I}_{RC,j}$)
            \State 2. Quality of Service Constraint ($\mathcal{I}_{QoSC,j}$)
            \State 3. Placement Cost Constraint ($\mathcal{I}_{PCC,j}$)
            \State 4. Scalability Overhead Constraint ($\mathcal{I}_{SOC,j}$)
            \State 5. Migration Overhead Constraint ($\mathcal{I}_{MOC,j}$)
            \If{placement does not meet constraints}
                \State{Discard placement}
            \EndIf
        \EndFor
        
         \State \textbf{2. Apply the target function to non-discarded placements}
        \ForAll{valid placements}
            \State Apply target function for placement \(j\):
            \State $F_j = \alpha \cdot QoS'_{PL_j} - \beta \cdot Cost'_{PL_j} - \gamma \cdot 
            OH'_{RO_j}$
            \State  It is based on the overall QoS, the placement costs, and the rescheduling operation overhead
        \EndFor
        
        \State \textbf{3. Select the best placement}
            \State $F_{\text{best}} = \max(F_j)$
            \State \(j_{best}\) is the placement that gets \(F_{best}\)
            
    \EndFunction
    \end{algorithmic}
    \end{algorithm}

     \begin{algorithm}[ht!]
    \renewcommand{\algorithmicrequire}{\textbf{Input:}}
    \renewcommand{\algorithmicensure}{\textbf{Output:}}
    \caption{Rescheduling}
    \label{alg:rescheduling}
    \begin{algorithmic}
    \Function{Check rescheduling}{ $j_{best}$, $j_{current}$, $\mathcal{I}_{RA, j_{best}}$}
    
        \Require $j_{best}$ \Comment{Best placement}
        \Require $j_{current}$ \Comment{Current placement}
        \Require $\mathcal{I}_{RA_{j_{best}}}$ \Comment{Resource Allocation Indicator}
        
        \Ensure $res\_op$ \Comment{Selected Rescheduling operation}
    
        \State \textbf{1. Determine rescheduling operation}
            \If{ $j_{best} = j_{current}$ and $\mathcal{I}_{RA, j_{best}} = \text{true}$ }
                \State $res\_op \gets \text{None}$
            \ElsIf{ $j_{best} = j_{current}$ and $\mathcal{I}_{RA, j_{best}} = \text{false}$ }
                \State $res\_op \gets \text{Scaling}$
            \ElsIf{ $j_{best} \neq j_{current}$ and $\mathcal{I}_{RA, j_{best}} = \text{true}$ }
                \State $res\_op \gets \text{Migration}$
            \ElsIf{ $j_{best} \neq j_{current}$ and $\mathcal{I}_{RA, j_{best}} = \text{false}$ }
                \State $res\_op \gets \text{Scaling and Migration}$
            \EndIf
             
        \State \textbf{2. Perform rescheduling}
        \State Execute $res\_op$ to transition to $j_{best}$
    \EndFunction
    \end{algorithmic}
    \end{algorithm}

    
    \item \textbf{Rescheduling}: Algorithm \ref{alg:rescheduling} selects the transition from \(j_{current}\) to \(j_{best}\). If both placements coincide, the service is maintained or scaled according to \(\mathcal{I}_{RA,j_{best}}\); otherwise, migration is triggered, combined with scaling when the destination allocation is insufficient.
    
    \item \textbf{Service execution}: The selected placement remains active until a context change triggers a new orchestration cycle.
   
\end{enumerate}

\section{Validation and Evaluation}
\label{validation}

This section presents a proof-of-concept evaluation of the proposed orchestration model across the ECC. First, we compare QoS-centric ($\uparrow \alpha$) and Cost-centric ($\uparrow \beta$) policies to illustrate how the decision-making logic adapts to different operational priorities. We then conduct a controlled ablation to isolate the contribution of XR-specific user differentiation to placement decisions under dynamic context changes, including user mobility, resource availability, and heterogeneous user requirements.

\subsection{Testbed}

The validation uses a deterministic, event-driven Python simulator operating at the orchestration level. The simulator maintains the active users and their profiles, user locations, node availability and physical capacities, current resource allocations, and the current service placement. After each context-change event, it updates this state, aggregates the resource demand of all active users, evaluates each available candidate node against the hard constraints, and computes its QoS, cost, rescheduling-overhead, and objective-function values. The highest-scoring feasible candidate is selected, the required action (no operation, scaling, migration, or both) is determined, and the placement and resource-allocation state are updated before processing the next event.

The general formulation supports placements comprising multiple infrastructure nodes. For this proof-of-concept validation, the XR service is instantiated as a monolithic service hosted at a single PoP ($PoP=1$), so each candidate placement $j$ corresponds to one infrastructure node $i$ ($j = i$). This configuration isolates the node-selection and contextual rescheduling logic under controlled conditions.

\begin{figure}[t!]
    \centering
    \includegraphics[width=1\linewidth]{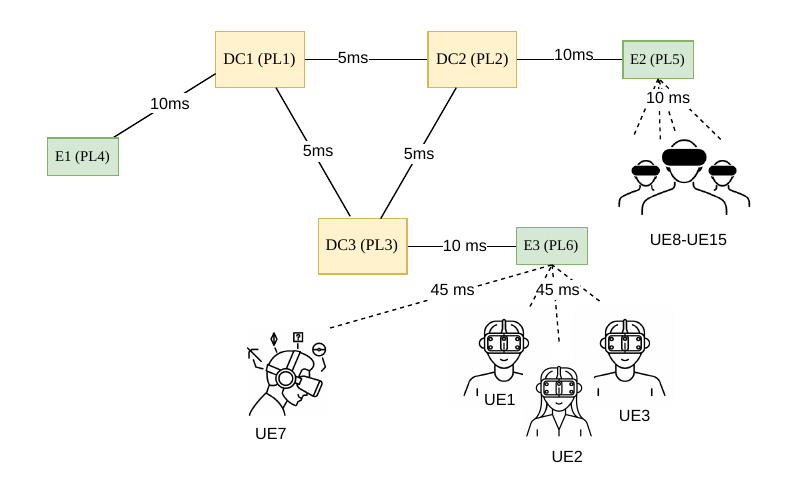}
    \caption{Network topology of the ECC testbed at Event 7.}
    \label{fig.topology}
\end{figure}

The environment comprises three distinct categories of nodes, each characterised by heterogeneous resource capabilities, as shown in Figure \ref{fig.topology}.

\begin{itemize}
    \item Cloud Nodes (C): High-capacity nodes with substantial resources, located at the cloud level. These nodes provide computational resources for hosting high-demand services. 
    \item Edge Nodes (E): Positioned closer to the users, and more resource-constrained. These play a crucial role in minimising user latency by processing data closer to their location.
    \item User Equipment (UE): UEs run lightweight XR clients and do not directly host the XR service (e.g., VR headsets, AR glasses).
\end{itemize}

Each node in the testbed is defined by its computational and resource capabilities. The specifications assigned to each node type for this validation purpose are described in Table \ref{tab:node_specs}. In order to keep the simulation realistic, node parameters are based on representative 5G equipment, while computational unit prices were obtained from Amazon Web Services \cite{aws_ec2_pricing}. The link latencies represent nominal one-way delays selected to emulate different proximity levels between cloud, edge, and user locations in a controlled orchestration-level evaluation. UEs only connect directly to Edge nodes. 

User concurrency is represented through the aggregate vCPU demand generated by all active users. For each candidate node $i$, this demand is compared with its physical capacity ($R_{max,i}$) and the resources currently allocated to the XR service ($R_{alloc,i}$). A candidate is infeasible when the aggregate demand exceeds ($R_{max,i}$); scaling is required when the demand exceeds ($R_{alloc,i}$) but remains within ($R_{max,i}$); otherwise, the existing allocation is sufficient. This abstraction captures service-level resource contention caused by concurrent users.

The model is exercised through a deterministic sequence of context-change events, including user arrivals with different requirements (Table \ref{tab:users}), user mobility, node-availability changes, and increasing resource demand. After each event, the complete candidate-evaluation and state-update process described above is repeated. This deterministic design enables each state transition and orchestration decision to be traced and reproduced step by step.

\begin{table}[t]
\renewcommand{\arraystretch}{1.4} 
\caption{Node types and specifications.}
\label{tab:node_specs}
\centering
\resizebox{\columnwidth}{!}{
\begin{tabular}{|l|l|l|l|l|l|}
\hline
Node Id & Type & $R_{max_i} (vCPU)$ & $R_{alloc_i} (vCPU)$ & $\varrho_i$ (per hour) & Lat to UEs \\ \hline

C1 & C & 
\begin{tabular}[c]{@{}l@{}}164 cores\end{tabular} & 
\begin{tabular}[c]{@{}l@{}}64 cores\end{tabular} &
\begin{tabular}[c]{@{}l@{}}0.0972/vCPU\end{tabular} &
- \\ \hline

C2 & C & 
\begin{tabular}[c]{@{}l@{}}164 cores\end{tabular} & 
\begin{tabular}[c]{@{}l@{}}64 cores\end{tabular} &
\begin{tabular}[c]{@{}l@{}}0.0972/vCPU\end{tabular} &
- \\ \hline

C3 & C & 
\begin{tabular}[c]{@{}l@{}}164 cores\end{tabular} & 
\begin{tabular}[c]{@{}l@{}}32 cores\end{tabular} &
\begin{tabular}[c]{@{}l@{}}0.0972/vCPU\end{tabular} &
- \\ \hline

E1 & E & 
\begin{tabular}[c]{@{}l@{}}32 cores\end{tabular} & 
\begin{tabular}[c]{@{}l@{}}6 cores\end{tabular} &
\begin{tabular}[c]{@{}l@{}}0.7/vCPU\end{tabular} &
2 ms \\ \hline

E2 & E & 
\begin{tabular}[c]{@{}l@{}}32 cores\end{tabular} & 
\begin{tabular}[c]{@{}l@{}}8 cores\end{tabular} &
\begin{tabular}[c]{@{}l@{}}0.7/vCPU\end{tabular} &
10 ms \\ \hline

E3 & E & 
\begin{tabular}[c]{@{}l@{}}16 cores\end{tabular} & 
\begin{tabular}[c]{@{}l@{}}8 cores\end{tabular} &
\begin{tabular}[c]{@{}l@{}}0.7/vCPU\end{tabular} &
45 ms \\ \hline

\end{tabular}
}
\end{table}

\begin{table}[t]
\renewcommand{\arraystretch}{1.3} 
\caption{User types and XR service specifications.}
\label{tab:users}
\centering
{
\begin{tabular}{|l|l|l|l|}
\hline
\( U \)       & UE1-UE6         & UE7                          & UE8-UE15     \\ \hline
\( R \)       & Participant     & Producer                     & Audience                    \\ \hline
\( I \)       & N-to-M          & 1-to-N                       & None                        \\ \hline
\( P_s \)     & PointCloud      & Avatar3D                     & None                        \\ \hline
\( L_{max} \) & 500 ms          & 500 ms                       & 500 ms                      \\ \hline
\( L_{proc} \)& 300 ms          & 100 ms                       & 70 ms                       \\ \hline
\( Rq \)  & Q1                  & Q2                           & Q3                          \\ \hline
\end{tabular}
}
\end{table}
\subsection{Model applicability}

\begin{table}[ht!]
\renewcommand{\arraystretch}{1.15}
\centering
\caption{Additional parameters and derived quantities used in the proof-of-concept validation.}
\label{tab:poc_parameters}
\scriptsize
\begin{tabularx}{\columnwidth}{|c|c|X|}
\hline
\textbf{Notation} & \textbf{Type} & \textbf{PoC instantiation} \\ \hline

$L_{max,u}$ & Input &
Maximum latency tolerated by each user profile (Table \ref{tab:users}). \\ \hline

$L_{proc,u}$ & Input &
Processing latency associated with each user profile (Table \ref{tab:users}). \\ \hline

$L_{net,u}$ & State &
Nominal network latency between user $u$ and candidate placement $j$. \\ \hline

$L_{mig,j}$ & State &
Nominal network latency between the current service location and candidate placement $j$. \\ \hline

$\varrho_i$ & Input &
Unit vCPU cost assigned to infrastructure node $i$ (Table \ref{tab:node_specs}). \\ \hline

$w_r,w_i,w_{r_q},w_{p_s}$ & Input &
Weights used to derive $UEL_u$; all set to 0.25. \\ \hline

$\omega_L,\omega_{Th}$ & Input &
QoS-component weights; $\omega_L=1$ and $\omega_{Th}=0$. \\ \hline

$\alpha,\beta,\gamma$ & Input &
Operator-defined weights for QoS, cost, and rescheduling overhead. \\ \hline

$\eta,\psi,\sigma$ & Input &
Overhead parameters instantiated as $\eta=1$, $\psi=1$, and $\sigma=0$. \\ \hline

$UEL_u$ & Derived &
User engagement level derived from $R$, $I$, $R_q$, and $P_s$.\\ \hline

$w_u$ & Derived &
User weight; instantiated as $w_u=UEL_u$. \\ \hline

$QoS_u$ & Derived &
Per-user QoS derived from processing and network latency. \\ \hline

$QoS'_{PL_j}$ & Derived &
Normalised aggregate QoS of candidate placement $j$. \\ \hline

$Cost_{PL_j},\,Cost'_{PL_j}$ & Derived &
Placement cost and its normalised value. \\ \hline

$S_{OH,j},\,M_{OH,j}$ & Derived &
Scaling and migration overheads associated with candidate $j$. \\ \hline

$OH_{RO,j},\,OH'_{RO,j}$ & Derived &
Total rescheduling overhead and its normalised value. \\ \hline

$L_{max}^{min}$ & Derived &
Minimum latency tolerance among the active users. \\ \hline

\begin{tabular}[c]{@{}c@{}}
$\mathcal{I}_{RC,j},\mathcal{I}_{RA,j},\mathcal{I}_{QoSC,j}$ \\
$\mathcal{I}_{PCC,j},\mathcal{I}_{SOC,j},\mathcal{I}_{MOC,j}$
\end{tabular}
& Derived &
Boolean indicators determining candidate-placement feasibility. \\ \hline

$F_j$ & Derived &
Multi-objective score used to rank feasible candidate placements. \\ \hline

\end{tabularx}
\end{table}
To apply the proposed orchestration framework to the ECC testbed, the general formulation is instantiated using the parameters available in the simulated environment. Table \ref{tab:poc_parameters} complements the general parameters marked in Table \ref{tab:xr_parameters} by summarising the additional configured inputs, state variables, and derived quantities used in the proof-of-concept. The orchestration process begins by evaluating the indicators associated with each constraint, as follows:

\begin{itemize}

    \item \textbf{RIC}: Equations \eqref{eq:irc_boolean} and \eqref{eq:ira_boolean} are evaluated by focusing exclusively on vCPU resources. $\mathcal{I}_{RC,j}$ verifies that the cumulative user demand ($\sum R_{usage,u}$) remains within the node's physical hardware limit ($R_{max_i}$). Simultaneously, $\mathcal{I}_{RA,j}$ assesses if the current logical vCPU allocation is sufficient.

    \item \textbf{QoSC}: The indicator $\mathcal{I}_{QoSC,j}$ (Equation \eqref{eq:qos_boolean}) is instantiated by focusing exclusively on end-to-end latency ($L_u$), assuming that throughput requirements are inherently satisfied by the network capacity. The total latency for each user is calculated as $L_u = L_{proc_u} + L_{net_u}$. While processing delay ($L_{proc_u}$) depends on the user role (Table \ref{tab:users}), network delay ($L_{net_u}$) is determined by the specific link between the UE and the candidate node. A placement is considered valid only if $L_u \leq L_{max_u}$ for all connected users ($L_{max_u}$ in Table \ref{tab:users}). This ensures that the orchestrator only selects nodes capable of maintaining the real-time interaction levels required for an immersive XR experience.

    \item \textbf{PCC}: We simplify the cost calculation by focusing on computational resources ($\varrho_i$), as energy consumption telemetry is not available in this testbed. Since the service is deployed as a single $PoP$, the total cost $Cost_{PL_j}$ is equal to $Cost_i$. We assume the operational budget $C_{opex}$ is sufficient for the evaluated scenarios; therefore, the indicator $\mathcal{I}_{PCC,j}$ (Equation \eqref{eq:pcc_boolean}) is consistently satisfied.
    
    \item \textbf{SOC}: We assume scaling-induced overhead ($S_{OH_j}$) remains within limits if physical capacity exists. Thus, $\mathcal{I}_{SOC,j}$, Equation \eqref{eq:soc_boolean}, is determined by the node's ability to accommodate $\Delta R_{usage}$. This is verified by checking if $S_g$ (the difference $R_{max_i} - R_{alloc_i}$) can absorb the new resource demand.

    \item \textbf{MOC}: The simulation assumes that migration overhead always satisfes the $M_{OH_{max}}$ threshold. Consequently, $\mathcal{I}_{MOC,j}$, Equation \eqref{eq:moc_boolean}, is always considered true.

\end{itemize}
Once the constraints are evaluated, the next step is to define the target function, Equation \eqref{eq:Fj}.

\begin{table}
\renewcommand{\arraystretch}{1.2} 
\caption{Scores for calculation of UEL.}
\label{tab:uel_scores}
\centering

\begin{tabular}{|l|l|l|}
\hline
\multirow{3}{*}{Role ($R$)}                     & Participant & 1  \\ \cline{2-3} 
                                                & Producer    & 0.5    \\ \cline{2-3} 
                                                & Audience    & 0.1  \\ \hline
\multirow{3}{*}{Interaction-type ($I$)}         & N-to-M      & 1    \\ \cline{2-3} 
                                                & 1-to-N      & 0.5  \\ \cline{2-3} 
                                                & None        & 0.1  \\ \hline
\multirow{3}{*}{Rendering quality profile ($Rq$)} & Q1        & 1    \\ \cline{2-3} 
                                                & Q2         & 0.5  \\ \cline{2-3} 
                                                & Q3         & 0.1  \\ \hline
\multirow{3}{*}{Self-perception ($Ps$)}         & PointCloud  & 1    \\ \cline{2-3} 
                                                & Avatar3D    & 0.5  \\ \cline{2-3} 
                                                & None        & 0.1  \\ \hline
\multirow{4}{*}{Parameter weights}              & $w_r$           & 0.25 \\ \cline{2-3} 
                                                & $w_i$           & 0.25 \\ \cline{2-3} 
                                                & $w_{r_q}$       & 0.25 \\ \cline{2-3} 
                                                & $w_{p_s}$       & 0.25 \\ \hline
\end{tabular}
\end{table}

\begin{table}
\renewcommand{\arraystretch}{1.2} 
\caption{Resources required for each perception type ($P_s$).}
\label{tab:req_resources}
\centering

\begin{tabular}{|l|l|l|l|}
\hline
   & PointCloud & Avatar3D & None \\ \hline

$R_{usage}(P_s) $ & \begin{tabular}[c]{@{}l@{}}
                    vCPU: 3 cores  
                \end{tabular} & 
                \begin{tabular}[c]{@{}l@{}}
                    vCPU: 2 cores  
                \end{tabular} & 
                \begin{tabular}[c]{@{}l@{}}
                    vCPU: 1 core  
                \end{tabular}\\ \hline
\end{tabular}
\end{table}

\begin{itemize}
    \item \textbf{Calculation of \(QoS'_{PL_j} \)}: The $QoS'_{PL_j}$ metric is instantiated by evaluating individual user performance weighted by their engagement level. Following Equations \eqref{Qos_pl_j} and \eqref{eq:wu}, we first determine the $UEL_u$, which reflects the intensity of the user's interaction. It is calculated as a weighted average of four parameters: Role ($R_u$), Interaction type ($I_u$), Rendering Quality Profile ($Rq_u$) and Self-perception type ($Ps_u$). Each parameter is assigned a score according to Tables \ref{tab:users} and \ref{tab:uel_scores}, resulting in the following expression:
    \begin{equation}
    UEL_u = w_{r_q} \cdot Rq_{u} + w_{r} \cdot R_u + w_{i} \cdot I_u + w_{p_s} \cdot Ps_u
    \end{equation}

    For this validation, the execution profile assumes full offloading for all candidate placements, so $UOL_j$ remains constant and does not differentiate them. User preferences ($FP_{up}$) are also uniform across the evaluated scenario. Consequently, the general weight $w_{u,j}$ is instantiated as $w_u = UEL_u$ allowing the proof-of-concept to isolate the effect of XR-specific user engagement on placement decisions. 
    
    
    To instantiate $QoS_u$, Equation \eqref{eq:qos_u} is simplified by focusing exclusively on the latency component ($\omega_L = 1, \omega_{Th} = 0$), including network and processing delays. Throughput requirements are assumed to be met by the network infrastructure. A clamping function is introduced to ensure the score remains within a valid range $[0, 1]$:
     \begin{equation}
        QoS_u=\max \left( 0, \; 1 - \frac{L_{proc_u} + L_{net_u}}{L_{\max_u}} \right)
    \end{equation}  
   
    The values for $L_{proc_u}$ and $L_{max_u}$ are predefined constants (Table \ref{tab:users}), while $L_{net_u}$ is a dynamic value determined by the network topology and the client's location. It represents the transmission delay of the network link between the UE and the candidate node $j$ being evaluated. With these values, we can calculate Equation \eqref{Qos_pl_j}. Finally, the value is normalised following Equation \eqref{eq:QoS_pl_normalised}.
    
    \item \textbf{Calculation of \(Cost'_{PL_j}\):} The operational cost is calculated based on the vCPU resources required by the users. Following the simplified version of Equations \eqref{eq:cost_total}, and \eqref{eq:costi}, we focus on the computational unit price ($\varrho_i$) of each node type, excluding energy consumption due to the lack of real-time telemetry:

    \begin{equation} 
    Cost_{PL_j}=Cost_{i} = \varrho_i \cdot \sum_{u \in U} R_{usage_u} 
    \end{equation}

    The specific vCPU demand for each user depends on their perception type ($P_s$), as detailed in Table \ref{tab:req_resources}. Since the service is deployed as a single $PoP$ in this testbed, the total placement cost $Cost_{PL_j}$ is equivalent to the individual node cost, $Cost_i$.

    For the normalisation process in Equation \eqref{eq:cost_pl_normalised}, the maximum possible cost $Cost_{PL_j}^{max}$ is also calculated considering only the unit prices. This ensures that the normalised metric $Cost'_{PL_j}$ remains consistent and accurately reflects the relative expense of each placement.

    \item \textbf{Calculation of \(OH'_{RO_j} \):} The rescheduling-overhead term provides an orchestration-level estimate of the temporal penalty associated with the actions required to reach each candidate placement.
    For the scaling overhead ($S_{OH,j}$), following Equation \eqref{eq:soh_formal}, we assume a constant actuation time ($T_{action}(scale)=0.1s$) representing the average delay to resize container resources. Following the sensitivity $\eta = 1$ established in the validation setup, if $\mathcal{I}_{RA}$ is false, $S_{OH,j}$ is defined by the physical duration of the task:
    \begin{equation}
    S_{OH,j} =
    \begin{cases}
    0.1s & \text{if } \mathcal{I}_{RA} = \text{false} \\
    0 & \text{otherwise}
    \end{cases}
    \end{equation}
    
    For migration overhead ($M_{OH,j}$), the actuation time is instantiated using the nominal latency between the current service location and candidate placement $j$, denoted by $L_{mig,j}$, as a topology-dependent proxy for the penalty of changing service location. Following Equation \eqref{eq:moh_formal}, we set $\psi = 1$ and $\sigma = 0$, focusing the proof-of-concept on the effect of infrastructure topology on rescheduling decisions.
    
    \begin{equation}
    M_{OH,j} =
    \begin{cases}
    L_{mig,j} & \text{if } j_{current} \neq j \\ 
    0 & \text{otherwise}
    \end{cases}
    \end{equation}
    
    Finally, the total rescheduling overhead $OH_{RO,j} = S_{OH,j} + M_{OH,j}$ (Equation \eqref{eq:oh_ro_j}) is normalised relative to the minimum latency tolerance among all connected users ($L_{max}^{min} = \min_{u \in U} L_{max,u}$). The normalised value $OH'_{RO,j}$ is capped at 1 to represent a total breach of the stability budget, so Equation \eqref{eq:oh_ro_normalised} is adjusted as:
    
    \begin{equation}
    \label{eq:oh_ro_normalised_final}
    OH'_{RO,j} = \min \left( 1, \frac{S_{OH,j} + M_{OH,j}}{L_{max}^{min}} \right)
    \end{equation}
    
    This ensures that $OH'_{RO,j} \in [0, 1]$, where a value of 1 indicates that the combined orchestration delay equals or exceeds the maximum tolerance of the most sensitive user, thereby incurring the maximum penalty in the objective function $F_j$.

 \end{itemize}
 


\subsection{Results and analysis}

The experimental sequence in Table~\ref{tab:scenario_events_final} evaluates the orchestrator under user arrivals, mobility, node availability changes, and edge saturation. The service is initially deployed at E3, where three participant users join. After E1 becomes available, these users move to E1 to exploit its lower latency, but the subsequent failure of E1 forces them back to E3. The scenario then introduces a producer and a large group of audience users, followed by additional participant users, progressively increasing the pressure on the edge infrastructure.

We evaluate the 10-event scenario under two representative policies:
\begin{itemize}
    \item QoS-centric: $\alpha=0.9, \beta=0.1, \gamma=0.1$
    \item Cost-centric: $\alpha=0.3, \beta=0.7, \gamma=0.1$
\end{itemize}

The weights $\alpha$, $\beta$, and $\gamma$ express operator priorities among feasible placements. Since hard constraints are evaluated first, changing these weights does not affect placement feasibility, but may alter the ranking of candidate placements, particularly when their objective scores are close. The QoS-centric and Cost-centric configurations are therefore used as representative policy profiles to illustrate this intended sensitivity. QoS-centric orchestration is more appropriate for highly interactive or mission-critical XR services, such as medical training or remote assistance, whereas Cost-centric orchestration may suffice for services with more relaxed QoS requirements or predominantly passive users, such as virtual tours.

 The top row in Figure~\ref{fig:results_tikz} shows $F_j$ prioritising experience through Edge infrastructure (73\% of the session). Following user mobility (Events 3 and 5), the engine triggers frequent rescheduling, including a joint Scaling and Migration (S\&M) at Event 3 to E1. Although this policy increases costs and $OH'_{RO}$ penalties, it maintains proximity to users. Finally, as E3 reaches its capacity in Event 7, the service migrates to the Cloud (C3) by Event 8 to ensure system scalability.

Conversely, the Cost-centric case prioritises infrastructure savings. An immediate migration from E3 to C3 (Event 1) minimises expenses, and the model avoids Edge nodes thereafter. Even when E1 becomes available, the orchestrator prefers C1 for its low-cost proximity. After returning to C3 in Event 5, the system remains static to minimise operational churn, prioritising long-term stability over minor latency gains even as additional audience users join the session.

\begin{table}[t]
\centering
\caption{Detailed Sequence of Events for XR Orchestration Test.}
\label{tab:scenario_events_final}
\small
\begin{tabularx}{\columnwidth}{clll}
\toprule
\textbf{Event} & \textbf{Type} & \textbf{Entity/Target} & \textbf{Description} \\ \midrule
0     & INIT & \{-\} & XR Service deployed at E3. \\
1     & CONN & \{UE1, UE2, UE3\}  & Participants join at E3. \\
2     & TOPO & E1 (ENABLE)       & Node E1 available. \\
3     & MOVI & UE1 \dots UE3 $\rightarrow$ E1 & Users move to E1. \\
4     & TOPO & E1 (DISABLE)      & Node E1 goes offline. \\
5     & MOVI & UE1 \dots UE3 $\rightarrow$ E3 & Users return to E3. \\
6     & CONN & \{UE7\}           & Producer joins at E3. \\
7     & CONN & \{UE8 \dots UE15\} & Audience join at E2. \\
8     & CONN & \{UE4\}           & Participant joins at E3. \\
9     & CONN & \{UE5\}           & Participant joins at E3. \\
10    & CONN & \{UE6\}           & Participant joins at E3. \\ \bottomrule
\end{tabularx}
\end{table}

The numerical trade-offs between these two strategies are summarised in Table \ref{tab:comparison_results}. The QoS-centric policy achieves a 4.26\% higher average $QoS'_{PL_j}$ (0.3239 versus 0.3106), whereas the Cost-centric policy reduces the average normalised placement cost by 41.39\% (from 0.1394 to 0.0817).

\begin{table}[t]
\renewcommand{\arraystretch}{1.4} 
\centering
\caption{Quantitative performance comparison between QoS-centric and Cost-centric orchestration policies (average values across 10 events)}
\label{tab:comparison_results}
\setlength{\tabcolsep}{2.8pt}
\begin{tabularx}{\columnwidth}{lccl}
\toprule
\textbf{Metric} & \textbf{QoS-centric (A)} & \textbf{Cost-centric (B)} & \textbf{Difference (\%)} \\ \midrule
Avg. $QoS'_{PL_j}$& 0.3239 & 0.3106 & 4.26\% (Gain in A) \\
Avg. $Cost'_{PL_j}$ & 0.1394 & 0.0817 & 41.39\% (Saving in B) \\
\bottomrule
\end{tabularx}
\end{table}


Finally, to assess the operational value of incorporating multi-user XR application context, we conduct a controlled ablation between an XR-context-aware configuration and a context-agnostic variant of the same orchestration model. Both configurations use the same topology, event sequence, constraints, cost model, rescheduling overhead model, objective-function structure, and QoS-centric policy. In the context-agnostic variant, the UEL-based differentiation is removed, so user role, interaction type, rendering profile, and perception mode do not influence user weighting during placement evaluation. Consequently, all users contribute uniformly to the aggregate QoS metric. This design isolates the effect of using XR application semantics to differentiate user requirements while keeping the remaining orchestration process unchanged.

Figure \ref{fig:results_tikz_2} shows how this contextual differentiation affects the orchestration outcome. The upper plot presents the service placement trajectories, while the lower plot reports the participant-specific $QoS'_{PL_j}$. Both configurations behave identically until users with different XR characteristics join at Events 6 and 7. During the congestion peak at Event 7, the context-agnostic configuration migrates the service to C2, reducing the participant-specific $QoS'_{PL_j}$ to 0.07. In contrast, the XR-context-aware configuration remains at E3 and achieves a participant-specific $QoS'_{PL_j}$ of 0.2163, representing a 208.97\% relative gain. This divergence shows that incorporating XR application semantics can materially influence placement decisions when heterogeneous users compete for constrained resources, allowing the configured policy to better preserve the QoS of prioritised users.

\begin{figure*}[t!]
\centering
\begin{tikzpicture}
\begin{groupplot}[
    group style={
        group size=2 by 3, 
        horizontal sep=2cm,
        vertical sep=1.2cm,
    },
    width=0.45\textwidth, 
    height=0.23\textwidth,
    xmin=0, xmax=11,
    xlabel={Event},
    grid=both,
    grid style={dashed, gray!30},
    xtick={0,1,2,3,4,5,6,7,8,9,10},
]


\nextgroupplot[
    ylabel={$F(j)$}, ymin=0.2, ymax=0.35,
    title={\small \textbf{QoS-Centric}},
    legend style={at={(1.1,1.3)}, anchor=south, legend columns=6}
]
\addplot[blue] coordinates {(1,0.2435)(2,0.2435)(3,0.3299)(4,0.3309)(5,0.2445)(6,0.2423)(7,0.2516)(8,0.2475)(9,0.2462)(10,0.2433)}; \addlegendentry{C1}
\addplot[cyan] coordinates {(1,0.2435)(2,0.2435)(3,0.3209)(4,0.3209)(5,0.2435)(6,0.2423)(7,0.2532)(8,0.2489)(9,0.2473)(10,0.2443)}; \addlegendentry{C2}
\addplot[teal] coordinates {(1,0.2535)(2,0.2535)(3,0.3219)(4,0.3209)(5,0.2525)(6,0.2523)(7,0.2599)(8,0.2562)(9,0.2551)(10,0.2523)}; \addlegendentry{C3}
\addplot[orange, thick] coordinates {(2,0.2162)(3,0.3386)(4,0.3436)(5,0.2212)}; \addlegendentry{E1}
\addplot[red, thick] coordinates {(1,0.2162)(2,0.2162)(3,0.2936)(4,0.2936)(5,0.2162)(6,0.2133)(7,0.2277)(8,0.2197)(9,0.2148)(10,0.2087)}; \addlegendentry{E2}
\addplot[magenta, thick] coordinates {(1,0.2662)(2,0.2662)(3,0.2986)(4,0.2936)(5,0.2612)(6,0.2633)(7,0.261)}; \addlegendentry{E3}

\nextgroupplot[
    ylabel={$F(j)$}, ymin=-0.2, ymax=0.1,
    title={\small \textbf{Cost-Centric}}
]
\addplot[blue] coordinates {(1,0.0426)(2,0.0446)(3,0.0734)(4,0.0744)(5,0.0456)(6,0.036)(7,0.0229)(8,0.0093)(9,-0.004)(10,-0.0171)}; 
\addplot[cyan] coordinates {(1,0.0426)(2,0.0446)(3,0.0704)(4,0.0704)(5,0.0446)(6,0.036)(7,0.0234)(8,0.0098)(9,-0.0036)(10,-0.0168)};
\addplot[teal] coordinates {(1,0.0466)(2,0.0486)(3,0.0714)(4,0.0704)(5,0.0476)(6,0.04)(7,0.0263)(8,0.0129)(9,-0.0003)(10,-0.0135)};
\addplot[orange, thick] coordinates {(2,-0.0147)(3,0.0261)(4,0.0271)(5,-0.0137)};
\addplot[red, thick] coordinates {(1,-0.0167)(2,-0.0147)(3,0.0111)(4,0.0111)(5,-0.0147)(6,-0.0346)(7,-0.0677)(8,-0.0989)(9,-0.1296)(10,-0.1601)};
\addplot[magenta, thick] coordinates {(1,0.0033)(2,0.0013)(3,0.0121)(4,0.0111)(5,0.0003)(6,-0.0186)(7,-0.0573)};


\nextgroupplot[
    ylabel={Action}, ymin=-0.5, ymax=3.5,
    ytick={0,1,2,3}, yticklabels={No-Op, Scal., Mig., S\&M},
    yticklabel style={rotate=45}
]
\addplot[only marks, thick, blue] coordinates {(1,1)(2,0)(3,3)(4,0)(5,2)(6,1)(7,1)(8,2)(9,0)(10,0)};

\nextgroupplot[
    ylabel={Action}, ymin=-0.5, ymax=3.5,
    ytick={0,1,2,3}, yticklabels={No-Op, Scal., Mig., S\&M},
    yticklabel style={rotate=45}
]
\addplot[only marks, thick, blue] coordinates {(1,2)(2,0)(3,2)(4,0)(5,2)(6,0)(7,0)(8,0)(9,0)(10,0)};

\nextgroupplot[
    ylabel={Placement}, ymin=-0.5, ymax=5.5,
    ytick={0,1,2,3,4,5}, yticklabels={C1, C2, C3, E1, E2, E3},
]
\addplot[thick, orange, mark=square*] coordinates {(0,5)(1,5)(2,5)(3,3)(4,3)(5,5)(6,5)(7,5)(8,2)(9,2)(10,2)};
\nextgroupplot[
    ylabel={Placement}, ymin=-0.5, ymax=5.5,
    ytick={0,1,2,3,4,5}, yticklabels={C1, C2, C3, E1, E2, E3},
]
\addplot[thick, orange, mark=square*] coordinates {(0,5)(1,2)(2,2)(3,0)(4,0)(5,2)(6,2)(7,2)(8,2)(9,2)(10,2)};\end{groupplot}
\end{tikzpicture}
\caption{Comparison of QoS-centric (left) and Cost-centric (right) policies. Rows depict the objective function $F(j)$, orchestration actions, and service placement trajectories across ten events.}
\label{fig:results_tikz}
\end{figure*}

\begin{figure}
\usepgfplotslibrary{groupplots}
\begin{tikzpicture}
\begin{groupplot}[
    group style={
        group size=1 by 2,
        vertical sep=1.2cm, 
    },
    width=0.45\textwidth,
    height=0.2\textwidth,
    xmin=0, xmax=11,
    xtick={1,2,3,4,5,6,7,8,9,10},
    grid=both,
    grid style={dashed, gray!30},
    xlabel={Event},
]

\nextgroupplot[
    ylabel={Placement},
    ymin=-0.5, ymax=5.5,
    ytick={0,1,2,3,4,5},
    yticklabels={C1, C2, C3, E1, E2, E3},
    legend pos=south west,
    legend style={font=\small},
]
\addplot[blue, thick, mark=*, mark options={fill=white}] 
    coordinates {(0,5)(1,5) (2,5) (3,3) (4,3) (5,5) (6,5) (7,1) (8,1) (9,1) (10,1)};
\addlegendentry{Context-agnostic}

\addplot[orange, thick, mark=square*] 
    coordinates {(0,5) (1,5) (2,5) (3,3) (4,3) (5,5) (6,5) (7,5) (8,2) (9,2) (10,2)};
\addlegendentry{XR-context-aware}
\draw[dashed, black!40] (axis cs:7,-0.5) -- (axis cs:7,5.5);

\nextgroupplot[
    ylabel={$QoS_{PL_{j}(Participants)}'$},
    ymin=0, ymax=0.45,
    legend pos=south west,
    legend style={font=\small},
]

\addplot[blue, thick, mark=*, mark options={fill=white}] 
    coordinates {(1,0.3100) (2,0.3100) (3,0.3960) (4,0.3960) (5,0.3100) (6,0.2325) (7,0.0700) (8,0.0862) (9,0.1000) (10,0.1120)};
\addlegendentry{Context-agnostic}

\addplot[red, thick, mark=square*] 
    coordinates {(1,0.3100) (2,0.3100) (3,0.3960) (4,0.3960) (5,0.3100) (6,0.2657) (7,0.2163) (8,0.2189) (9,0.2302) (10,0.2384)};
\addlegendentry{XR-context-aware}

\draw[<->, >=stealth, thick, black!70] (axis cs:7,0.0700) -- (axis cs:7,0.2163) 
    node[midway, right] {+208.97\%};
\draw[dashed, black!40] (axis cs:7,0) -- (axis cs:7,0.45);

\end{groupplot}
\end{tikzpicture}
\caption{Comparison of service placement (top) and participant-specific $QoS_{PL_{j}}$ (bottom) for Context-agnostic and XR-context-aware orchestration.}
\label{fig:results_tikz_2}
\end{figure}

\section{Conclusions and future work}
\label{conclusions}

This paper presents a context-aware orchestration model for multi-user XR services across the ECC. The work addresses a specific gap identified in the analysed literature: although current solutions provide general monitoring, scaling, migration, and QoS/QoE management capabilities, they lack a unified and configurable representation of the internal characteristics of multi-user XR experiences that can directly inform orchestration decisions. To address this gap, the proposed model parametrises XR-relevant context across four architectural layers and formalises its integration into placement and rescheduling decisions. The User Engagement Level (UEL) and User Offloading Level (UOL) provide analytical abstractions for connecting high-level service characteristics with infrastructure-level evaluation.

The proof-of-concept validation illustrates two relevant properties of the model. First, the QoS-centric and Cost-centric configurations show how operator-defined policies can adapt orchestration decisions to different performance and economic priorities. In the evaluated scenario, the Cost-centric policy achieves substantial cost savings with only a limited reduction in aggregate QoS. Second, the controlled ablation shows that incorporating XR application semantics can materially change the selected placement under resource contention and help preserve the QoS of users prioritised by the service policy. Together, these results demonstrate the value of introducing configurable multi-user XR context into ECC orchestration.

Future work will focus on deploying the model over a real and complex test environment, involving a federated infrastructure composed of edge and cloud domains with varying latency, capacity, and sustainability characteristics. This will enable a more thorough evaluation of the scalability, interoperability, and responsiveness of the model under real-world constraints, including mobility, multi-tenancy, and network variability.

\section*{Acknowledgment}
This research was supported by the SNS-JU Horizon Europe Research and Innovation programme, under Grant Agreement 101096838 for 6G-XR project. It has also been partially funded by the Spanish National Plan for Scientific and Technical Research and Innovation, under award number CER-20231018, project 6GDiferente and by the national project Grant PID2022-137329OB-C44 funded by MICIU/AEI/10.13039/501100011033 titled ”EnablIng Native-AI Secure deterministic 6G networks for hyPer-connected envIRonmEnts” (6G-INSPIRE).


 
%

\bibliographystyle{IEEEtran}
\bibliography{2-main.bib}











 
\vspace{-33pt}
\begin{IEEEbiography}[{\includegraphics[width=1in,height=1.25in,clip,keepaspectratio]{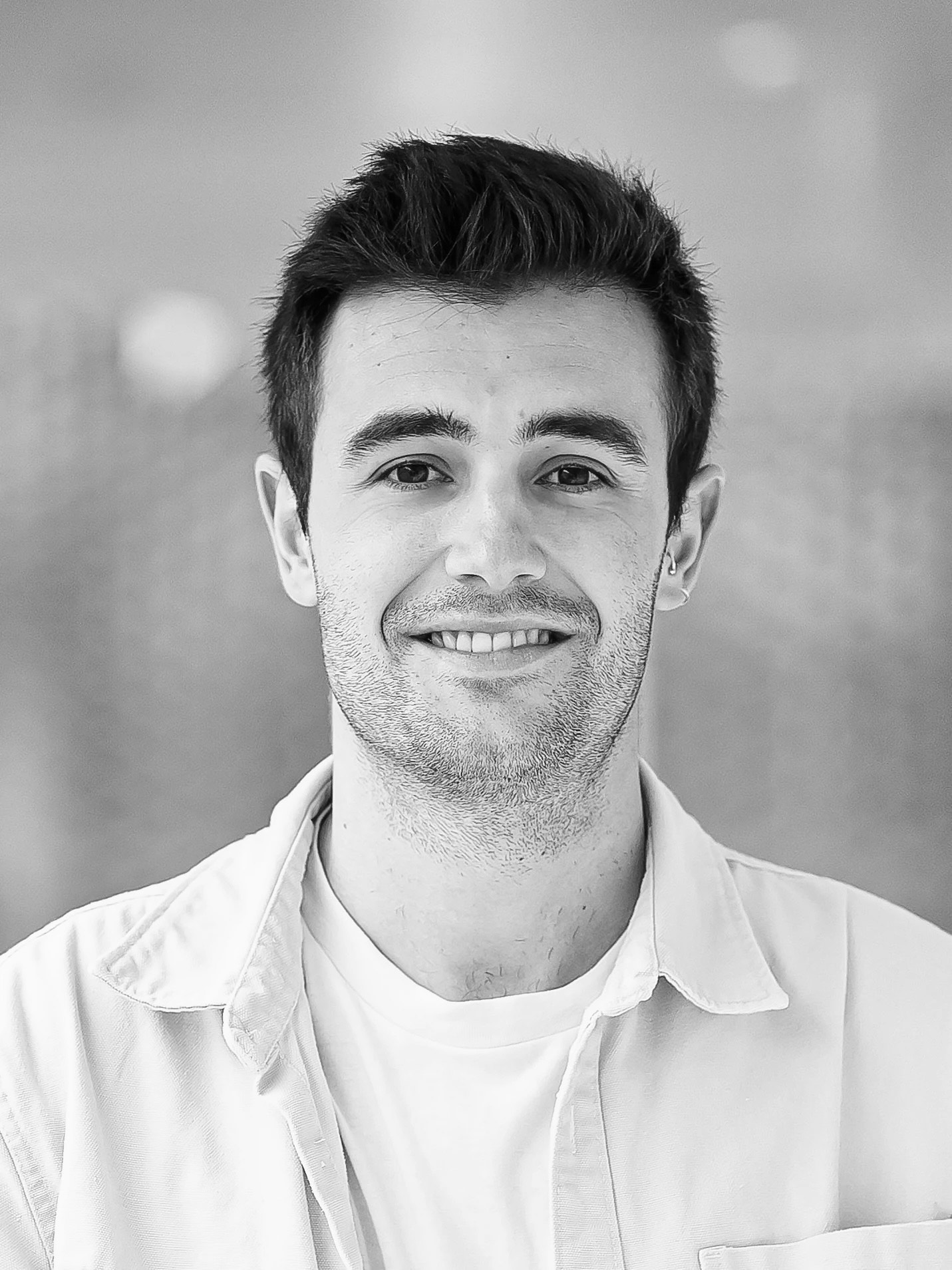}}]{Inhar Yeregui} received the B.Sc. and M.Sc. degrees in telecommunications systems from the University of the Basque Country, Spain in 2020 and 2022, respectively. He is currently working towards the Ph.D. degree with the University of the Basque Country in collaboration with the Digital Media and Communications Department of Vicomtech Foundation, San Sebastian, Spain, in projects dealing with 5G communications, virtualisation paradigm, and cloud-native XR services.
\end{IEEEbiography}

\vspace{-33pt}
\begin{IEEEbiography}[{\includegraphics[width=1in,height=1.25in,clip,keepaspectratio]{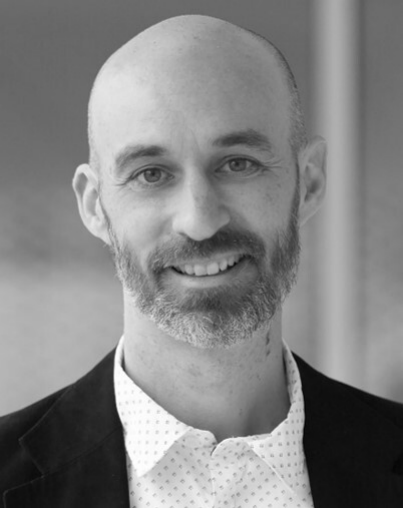}}]{\'{A}ngel Mart\'{i}n} is with the Department of Digital Media, Vicomtech. He received his PhD degree (2018) from UPV/EHU and his engineering degree (2003) from University Carlos III. He worked in media streaming and encoding research at Prodys (2003-2005) and Telefonica (2005-2008). Then, he worked in the field of ubiquitous and pervasive computing at Innovalia (2008-2010). Currently, he is at Vicomtech, working in multimedia services and 5G infrastructures projects.
\end{IEEEbiography}

\vspace{-33pt}
\begin{IEEEbiography}[{\includegraphics[width=1in,height=1.25in,clip,keepaspectratio]{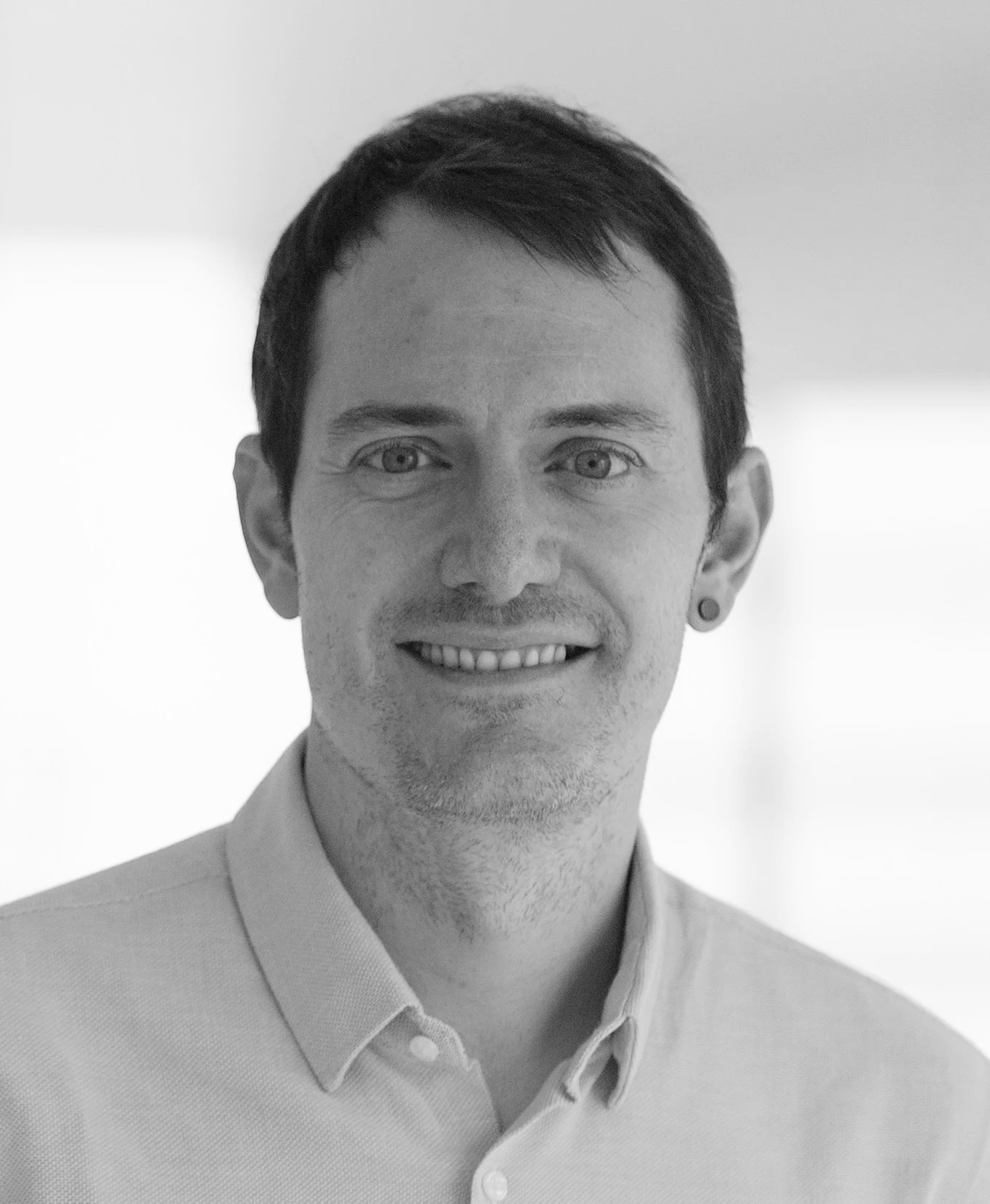}}]{Mikel Zorrilla} received the telecommunications engineering degree from the School of Engineering, Mondragon Unibertsitatea, Mondragon, Spain, in 2007, advanced degree in computer science and the Ph.D. degree from the University of the Basque Country, Spain, in 2012 and 2016, respectively. His Ph.D. dissertation was entitled “Interoperable Technologies for Multi-device Media Services” . From 2002 to 2006, he was a Researcher with Ikerlan S. Coop. and with Vicomtech since 2007. He has been an Associate Professor of multimedia technologies with Deusto University, Bilbao, Spain, in 2014. Since 2018, he has been an Associate Professor with the Universitat Oberta de Catalunya, Barcelona, Spain, in the multimedia field. He is currently the Director of the Digital Media and Communications Department, Vicomtech, San Sebastian, Spain. He has led and participated in several national and international research and innovation projects. One of them is TRACTION, a European H2020 project where opera co-creation had been explored as a path for social and cultural inclusion. His main research interests include interactive media technologies, media standards, video workflow management, and 5G technologies.
\end{IEEEbiography}

\vspace{-33pt}
\begin{IEEEbiography}[{\includegraphics[width=1in,height=1.25in,clip,keepaspectratio]{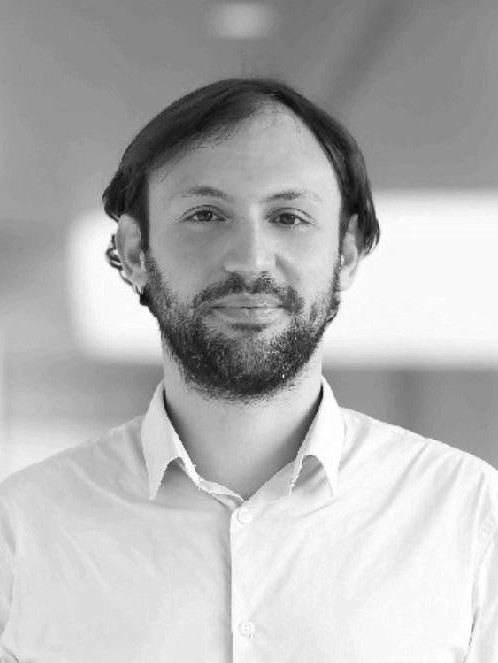}}]{Roberto Viola} is with the Department of Digital Media and Communications, Vicomtech. He received his advanced degree in Telecommunication Engineering in 2016 from University of Cassino and Southern Lazio and his PhD degree in 2021 from University of the Basque Country. He is Senior Researcher at Vicomtech, where he is involved in several projects concerning multimedia services and network infrastructures.
\end{IEEEbiography}

\vspace{9pt}
\begin{IEEEbiography}[{\includegraphics[width=1in,height=1.25in,clip,keepaspectratio]{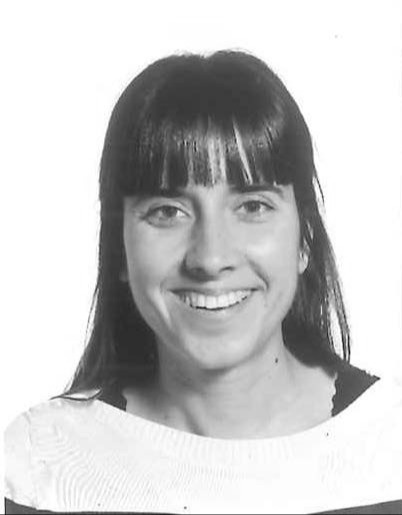}}]{Jasone Astorga}

    received the B.Sc. and M.Sc. degrees in Telecommunication Engineering and the Ph.D. degree from the University of the Basque Country (UPV/EHU), in 2004 and 2013, respectively. From 2004 to 2007, she worked with Nextel S.A., a telecommunications enterprise. She joined the UPV/EHU in 2007 as a lecturer and a researcher with the I2T Research Laboratory. She is currently an Assistant Professor with the UPV/EHU. She has led and contributed to numerous local, national, and European research projects in cybersecurity for industrial environments, Software Defined Networking (SDN), Network Function Virtualisation (NFV), and IoT security. She has co-authored numerous journal papers and conference contributions, and has supervised four PhD candidates and numerous master's and bachelor's projects.
\end{IEEEbiography}

\vspace{9pt}
\begin{IEEEbiography}[{\includegraphics[width=1in,height=1.25in,clip,keepaspectratio]{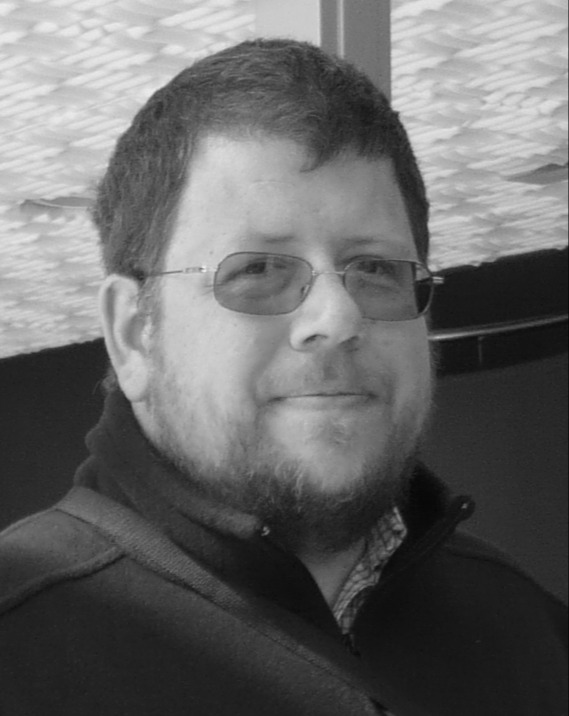}}]{Eduardo Jacob}

\{Senior Member, IEEE\} received the M.S. in Industrial Engineering and Ph.D. degree in Information and Communication Technologies in 2001 from the University of the Basque Country (UPV/EHU), Spain. He is currently Full Professor with the Department of Communications Engineering at UPV/EHU. He is the I2T Research Laboratory principal investigator and SmartQuanT4E and SmartNets4E research infrastructures director. His research interests include computer networks, network security, and advanced network architectures, with a focus on quantum-safe and time-sensitive networking. He has authored and co-authored numerous publications in international journals and conferences, and actively participates and lead national and European research projects.
\end{IEEEbiography}


 \vfill

\end{document}